\documentclass[twocolumn]{aastex63}

\usepackage{multirow}
\usepackage{amsmath}
\usepackage{graphicx}
\usepackage{float}
\usepackage{xcolor}

\shorttitle{Mapping the 3D Kinematical Structure of the Gas Disk of HD 169142}
\shortauthors{Yu et al.}

\graphicspath{{./}{figures/}}

\newcommand{\PID}[1]{[\href{https://almascience.nrao.edu/aq/?project_code/#1}{#1}]}

\begin{document}

\title{Mapping the 3D Kinematical Structure of the Gas Disk of HD 169142}

\author[0000-0002-0971-6078]{Haochuan Yu}
\affiliation{Department of Astronomy, Beijing Normal University, Beijing 100875, People's Republic of China}
\affiliation{Department of Physics, University of Oxford, Oxford OX1 3RH, UK}

\author[0000-0003-1534-5186]{Richard Teague}
\affiliation{Center for Astrophysics \textbar{} Harvard \& Smithsonian, 60 Garden Street, Cambridge, MA 02138, USA}

\author[0000-0001-7258-770X]{Jaehan Bae}
\altaffiliation{NASA Hubble Fellowship Program Sagan Fellow}
\affiliation{Department of Astronomy, University of Florida, Gainesville, FL 32611, USA}
\affiliation{Earth and Planets Laboratory, Carnegie Institution for Science, 5241 Broad Branch Road NW, Washington, DC 20015, USA}

\author[0000-0001-8798-1347]{Karin \"{O}berg}
\affiliation{Center for Astrophysics \textbar{} Harvard \& Smithsonian, 60 Garden Street, Cambridge, MA 02138, USA}

\begin{abstract}
The disk around HD~169142 has been suggested to host multiple embedded planets due to the range of structures observed in the dust distributions. We analyze archival ALMA observations of $^{12} \mathrm{CO \ (2-1)}$, $^{13} \mathrm{CO \ (2-1)}$, and $\mathrm{C}^{18} \mathrm{O \ (2-1)}$ to search for large-scale kinematic structures associated with other embedded planets in the outer disk. At 125~au, we identify a coherent flow from the disk surface to the midplane, traced by all three CO isotopologues, and interpret it as a meridional flow, potentially driven by an embedded planet. We use changes in the rotation speed of the gas to characterize the physical structure across this region, finding that at 125~au the CO emission traces regions of increased gas pressure, despite being at a surface density minimum. Developing a simple analytical model, we demonstrate that the physical structure of the gap can have non-trivial responses to changes in the surface density, consistent both with previous thermo-chemical models, and the conditions inferred observationally. Applying this technique to a range of sources will allow us to directly confront theoretical models of gap-opening in protoplanetary disks.
\end{abstract}


\section{Introduction}
\label{sec:intro}

With the Atacama Large (sub-) Millimetre Array (ALMA), we have the ability to characterize the physical and chemical structure of protoplanetary disks, the formation location of exoplanets. Particularly, spatially and spectrally resolved molecular line emission enables us to study the gas structure of the disk.

The dynamical structure of the disk encodes information about any potential planet-disk interactions, as embedded planets are predicted to drive large perturbations in the gas velocity structure during their formation (e.g., \citealt{2015ApJ...811L...5P,2018MNRAS.480L..12P,2019NatAs...3.1109P,2020arXiv200904345D,2021ApJ...912...56B}). The line-of-sight velocity, $v_{\mathrm{LOS}}$ of the gas, is a combination of the systemic velocity, $v_{\mathrm{LSR}}$, and the internal motions of the disk, including the rotational velocity, $v_{\mathrm{rot}}$, radial velocity, $v_{\mathrm{rad}}$ and vertical velocity, $v_{\mathrm{z}}$, given by 

\begin{equation} \label{eq:vlos}
v_{\mathrm{LOS}}=v_{\mathrm{rot}} \sin (i) \cos (\phi)+v_{\mathrm{rad}} \sin (i) \sin (\phi)+v_{\mathrm{z}} \cos (i)+v_{\mathrm{LSR}},
\end{equation}

\noindent where $\phi$ is the azimuthal angle in the frame of reference of the disk, $i$ is the inclination of the disk. If we assume that the disk is azimuthally symmetric, then we can disentangle the velocity structure of the disk according to their different dependence on $\phi$ \citep[e.g.,][]{2019Natur.574..378T, 2020arXiv200904345D}. 

As protoplanetary disks are expected to be in hydrostatic equilibrium, the rotational velocity, $v_{\mathrm{rot}}$, is closely linked to the physical structure of the disk \citep[e.g.,][]{2012ApJ...757..129R}, through

\begin{equation} \label{eq:4}
\frac{v_{\mathrm{rot}}^{2}}{r}=\frac{G M_{\mathrm{star}} r}{\left(r^{2}+z^{2}\right)^{3 / 2}} +\frac{1}{\rho_{\mathrm{gas}}} \frac{\partial P}{\partial r}\mathrm{,}
\end{equation}

\noindent where $P=n_{\mathrm{gas}} k T$ is the gas pressure, $\rho_{\mathrm{gas}}=\mu_{\mathrm{gas}} n_{\mathrm{gas}}$ is the gas density, $\mu_{\mathrm{gas}}$ is the mean molecular mass of the gas. As embedded planets are predicted to open large gaps in the gas surface density \citep[e.g.,][]{1984ApJ...285..818P}, leading to large radial changes in the gas pressure gradient, we therefore expect to see variations in the rotation speed of the gas if there are massive enough embedded planets \citep[][]{2015MNRAS.448..994K}. Thermo-chemical modeling by \citet{2020A&A...642A.165R} showed that this variation in pressure arises to changes in both the gas density and temperature.

The radial velocity, $v_{\mathrm {rad}}$, and vertical velocity, $v_{\mathrm z}$, depend on any internal, large scale gas flows within the disk. One of such flow is the meridional flow, where the disk viscously spreads to `fill in' large gaps in the gas surface density, including associated with embedded planets  \citep{2014Icar..232..266M, 2014ApJ...782...65S, 2016ApJ...832..105F}. Using $^{12}$CO observations of the disk around HD~163296, \citet{2019Natur.574..378T} detected meridional flows at the radial location of two gaps in the distributions of mm-sized grains and at the radial location of a previously inferred planet \citep{2018ApJ...860L..13P}, arguing that a planetary origin was the most likely scenario.

In this work, we study the velocity structure of the disk of HD~169142, a Herbig Ae star \citep{1994A&AS..104..315T} with a stellar mass of $1.65 \pm 0.05 \ {M}_{\odot}$ \citep{2018A&A...614A.106C}, an age of ${\sim} 10~\mathrm{Myr}$ \citep{2017ApJ...850...52P} and at a distance of $d = 113.6 \pm 0.8~\mathrm{pc}$ \footnote{We note that with Gaia EDR3, the distance has been updated to 114.9 pc \citep{2021A&A...649A...1G}.This change in distance of $\approx 1$~pc will result in a change of only $\approx 1\%$ of the derived stellar mass.} \citep{2018AJ....156...58B, 2018A&A...616A...1G}. The disk is observed almost at a face-on orientation, with an inclination $i = 13\degr$ \citep{2006AJ....131.2290R}, and with a position angle of ${\rm PA} = 5\degr$ \citep{2017A&A...600A..72F}.

This source is of particular interest as both millimeter continuum and near-infrared (NIR) observations, tracing mm sized grains and sub-\micron{} grains, respectively, have revealed a variety of structures (e.g., \citealt{2013ApJ...766L...2Q, 2014ApJ...792L..23R, 2014ApJ...792L..22B, 2017ApJ...838...20M, 2018MNRAS.473.1774L, 2017ApJ...838...97M, 2019ApJ...881..159M}) in the disk dust distribution indicative of embedded planets. Two major ring-like structures, located at radii ${\sim} 0\farcs22$ (${\sim} 25~\mathrm{au}$; B25) and ${\sim} 0\farcs53$ (${\sim} 60~\mathrm{au}$; B60), have been identified by both millimeter continuum and near-IR observations (e.g., \citealt{2014ApJ...791L..36O, 2017A&A...600A..72F, 2017ApJ...838...97M, 2018MNRAS.474.5105B, 2018A&A...614A.106C, 2019AJ....158...15P, 2019ApJ...881..159M}). With higher angular resolution observations, \citet{2019AJ....158...15P} observed that B60 was actually composed of three narrow rings. Two dust-depleted regions were observed: an inner cavity ($r \lesssim 20~\mathrm{au}$; \citealt{2017A&A...600A..72F}; D10) and a gap centered at ${\approx} 41$~au \cite[][D41]{2017A&A...600A..72F} bounded by B25 and B60. In terms of the gas distribution, \citet{2017A&A...600A..72F} observed depletions of CO isotopologue emission in D41, but not in D10, attributing them to drops in CO column density. The authors were unable to distinguish whether these CO column density drops reflect a gas surface density depletion or a CO abundance depletion.

One possible scenario for explaining the creation of these rings and gaps is planet–disk interactions \citep{2017A&A...600A..72F, 2018MNRAS.474.5105B, 2019ApJ...881..159M}. Given that the density drops of CO isotopologues are only observed in D41, not in D10, and the lack of azimuthal asymmetries in the two dust rings, B25 and B60, \citet{2017A&A...600A..72F} inferred that two planets of masses of ${\sim} 0.1-1~M_{\rm Jup}$ and ${\sim} 1-10 \ M_{\rm Jup}$ in D10 and D41, respectively, are needed to explain the observed surface density depletions. More recent hydrodynamic simulations published by \citet{2018MNRAS.474.5105B} suggested that one planet might not be sufficient to open up a gap wide enough as D41 (width ${\sim} 20$~au), arguing that two $1~M_{\mathrm {Jup}}$ planets would better reproduce the observations. Besides, hydrodynamic simulations by \citet{2020ApJ...888L...4T} were able to reproduce observed long-lived dust rings between two giant planets, with masses of ${\sim} 2.4~M_{\rm {Jup}}$ at 17.4 and 50~au. In the outer regions of the disk, \citet{2019AJ....158...15P} found the triple-ring structure they observed might relate to the formation of an embedded mini-Neptune sized planet at $\sim$ 65~au based on their hydrodynamic simulations.

Besides rings and gaps, other structures have been observed, suggestive of embedded planets. A blob was observed with NIR scattered light observations at 36.4~au, with a rotation speed of $-2.08 \pm 0.25\degr \ \mathrm{yr}^{-1}$, which might relate to a $2.2_{-0.9}^{+1.4} \ M_{\rm Jup}$ still-accreting planet at about 38~au. This plane could also be responsible for carving the D41 and exciting a spiral-like structure observed within B25 and B60 \citep{2019A&A...623A.140G}. 

In \citet{2020arXiv200711565B}, a time-variable surface brightness dip was found spanning several observations covering 6 years. The dip is located in B25, with a rotation speed of ${\sim} 6\degr~\mathrm{yr}^{-1}$. Based on the width of the shadow, \citet{2020arXiv200711565B} suggested that the dip could be due to a $\sim 1-10~M_{\rm Jup}$ planet at 12~au.

Given that there is strong evidence for multiple embedded planets in the inner disk around HD~169142, we explore the dynamic structure of the disk to search for kinematic evidence of embedded planets in the outer disk (i.e. $\geq$ 80~au), reginos that are inaccessible when studying the dust distributions. We measured three-dimensional velocity structures of different layers of the gas in this disk, traced by $^{12}\mathrm{CO \ (2-1)}$, $^{13} \mathrm{CO \ (2-1)}$, and $\mathrm{C}^{18} \mathrm{O \ (2-1)}$ emission, which is described in Section \ref{sec:measuring v}. In Section \ref{sec:model}, we use an analytical disk model to estimate the perturbations in the gas density based on changes in the gas rotational velocity. We discuss the findings in Section \ref{sec:discussion}, and summarize in Section \ref{sec:conclusion}.

\section{Observations}
\label{sec:observations}

We combine observations from three archival ALMA projects, with details given in Table~\ref{tab:observations}. These observations all target the $(2-1)$ transition of the three main CO isotopologues, $^{12}$CO, $^{13}$CO and C$^{18}$O. Each archival project was calibrated using the scripts provided by the ALMA Archive and the required version of \texttt{CASA} \citep{McMullin_ea_2007}, before transferring to \texttt{CASA v5.8.0} for self-calibration and imaging.

\begin{deluxetable*}{ccccccccc}
\tablecaption{Summary of Archival Observations \label{tab:observations}}
\tablehead{Project I.D. & P.I. & Date & On-Source & $N_{\rm ant}$ & Baselines & & Calibrators & \\
 & & & (min) & & (m) & (flux) & (bandpass) & (phase)} 
\startdata
\PID{2013.1.00592.S} & D. Fedele    & 30th Aug. 2015 & 32.7 & 35 & 15 -- 1466 & Ceres      & J1924-2914 & J1812-2836 \\\hline
                     &              & 14th Sep. 2016 & 19.8 & 38 & 15 -- 3248 & J1733-1304 & J1924-2914 & J1820-2528 \\
\PID{2015.1.00490.S} & M. Honda     & 14th Sep. 2016 & 19.8 & 38 & 15 -- 3248 & J1924-2914 & J1924-2914 & J1820-2528 \\
                     &              & 14th Sep. 2016 & 19.8 & 38 & 15 -- 3248 & J1924-2914 & J1924-2914 & J1820-2528 \\\hline
\PID{2015.1.01301.S} & J. Hashimoto & 17th Sep. 2016 & 19.8 & 40 & 15 -- 3144 & J1924-2914 & J1924-2914 & J1820-2528 \\
\enddata
\end{deluxetable*}

\subsection{Self Calibration and Imaging}

Each execution block was first imaged, then the phase center measured using the \texttt{imfit} task in \texttt{CASA}. These phase centers allowed all execution blocks to be centered using the \texttt{fixvis} command, before updating a common phase center for all execution blocks using the \texttt{fixplanets} command to account for any proper motion of the source. Then, each project was self-calibrated individually, following the procedure described in \citet{Andrews_ea_2018}. Only continuum base bands and line-free channels were considered. Numerous phase solutions were used, starting at infinity and decreasing to 10s, or stopping sooner when the signal to noise of the data did not improve. A single round of amplitude self calibration was performed for each project with an infinite solution interval. The self calibration improve the signal-to-noise ratio of the continuum by factors of 2, 3 and 4 for the 2013.1.00592, 2015.1.00490 and 2015.1.01301.S projects, respectively. The solutions were applied to the spectral line data, before the continuum being subtracted using the \texttt{uvcontsub} task.

Imaging of the continuum and spectral line data combined all three projects. Briggs weighting was chosen with a robust value of 0.5 which was found to give a good trade off between spatial resolution and sensitivity, yielding a synthesized beam size of $0\farcs145 \times 0\farcs089$ ($53\fdg5$) for the continuum, $0\farcs182 \times 0\farcs135$ ($60\fdg4$) for $^{12}$CO, $0\farcs190 \times 0\farcs140$ ($61\fdg1$) for $^{13}$CO, and $0\farcs191 \times 0\farcs146$ ($57\fdg8$) for C$^{18}$O. The channel spacing was set to $165~{\rm m\,s}^{-1}$ due to the coarser spectral resolution of the 2015.1.01301.S data. The images were masked using a Keplerian mask\footnote{\url{https://github.com/richteague/keplerian_mask}} assuming literature values, including stellar mass, inclination, and position angle for the system. The mask was enlarged through convolution of a 1\arcsec{} circular kernel to ensure all emission was covered by the mask. Images were made using the `multi-scale' algorithm adopting pixel scales of (0, 6, 12, 18, 24), and were CLEANed down to a $3\sigma$ level, where $\sigma$ was the RMS measured in a line free channel. The RMS is $52.9~\mu {\rm Jy\,beam^{-1}}$ for the continuum, $1.1~{\rm mJy\,beam^{-1}}$ for both $^{12}$CO and $^{13}$CO, and $0.8~{\rm mJy\,beam^{-1}}$ for C$^{18}$O. To account for the mismatches in the flux units of the model images and the residuals, we apply the `JvM' correction proposed in \citet[Appendix A]{Jorsater_vanMoorsel_1995} and described in more detail in \citet{2021arXiv210906188C}. For the continuum data, we adopt the same imaging properties, and clean down to $3\sigma$ where $\sigma$ was measured from continuum-free regions offset from the center.

\subsection{Results}
\begin{figure*}
    \centering
    \includegraphics[width=0.9\textwidth]{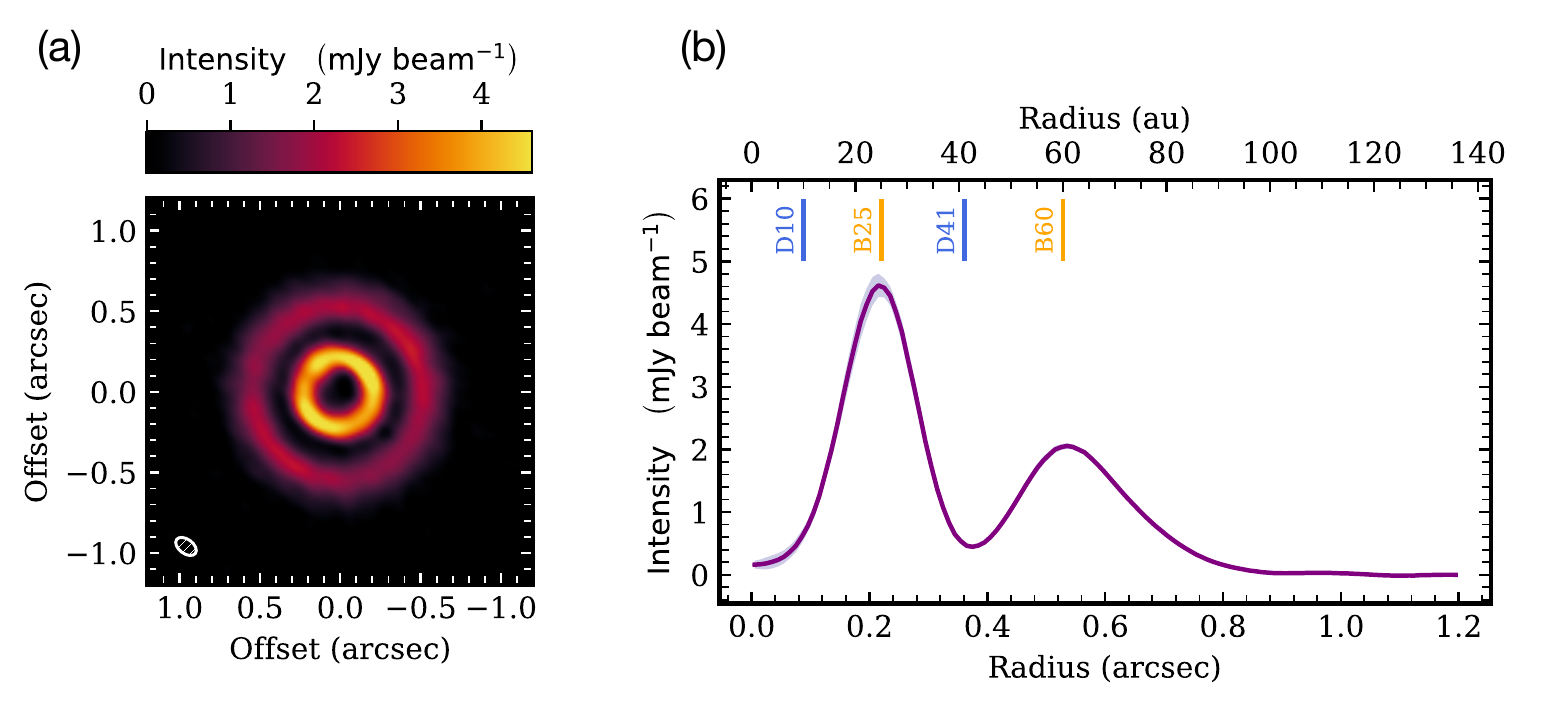}
    \caption{An overview of the dust observation data. Panel (a) shows the continuum map, while the panel (b) shows its radial profile. The two major rings, B25 and B60, and the two major regions of dust-depletion, D10 and D41, found in previous studies are labeled in panel (b). The shaded region in the radial profile shows the azimuthal scatter in each radial bin.}
    \label{fig:ov-dust}
\end{figure*}

We show continuum map in Figure~\ref{fig:ov-dust}a, and calculate its azimuthally averaged radial profile with \texttt{GoFish}\footnote{\url{https://github.com/richteague/gofish}} \citep{GoFish} package, shown in Figure~\ref{fig:ov-dust}b. The shaded region in the radial profile shows the azimuthal scatter in each radial bin. We label the two major rings, B25 and B60, and the two major regions of dust-depletion, D10 and D41, in the radial profile.

\begin{figure*}
    \centering
    \includegraphics[width=\textwidth]{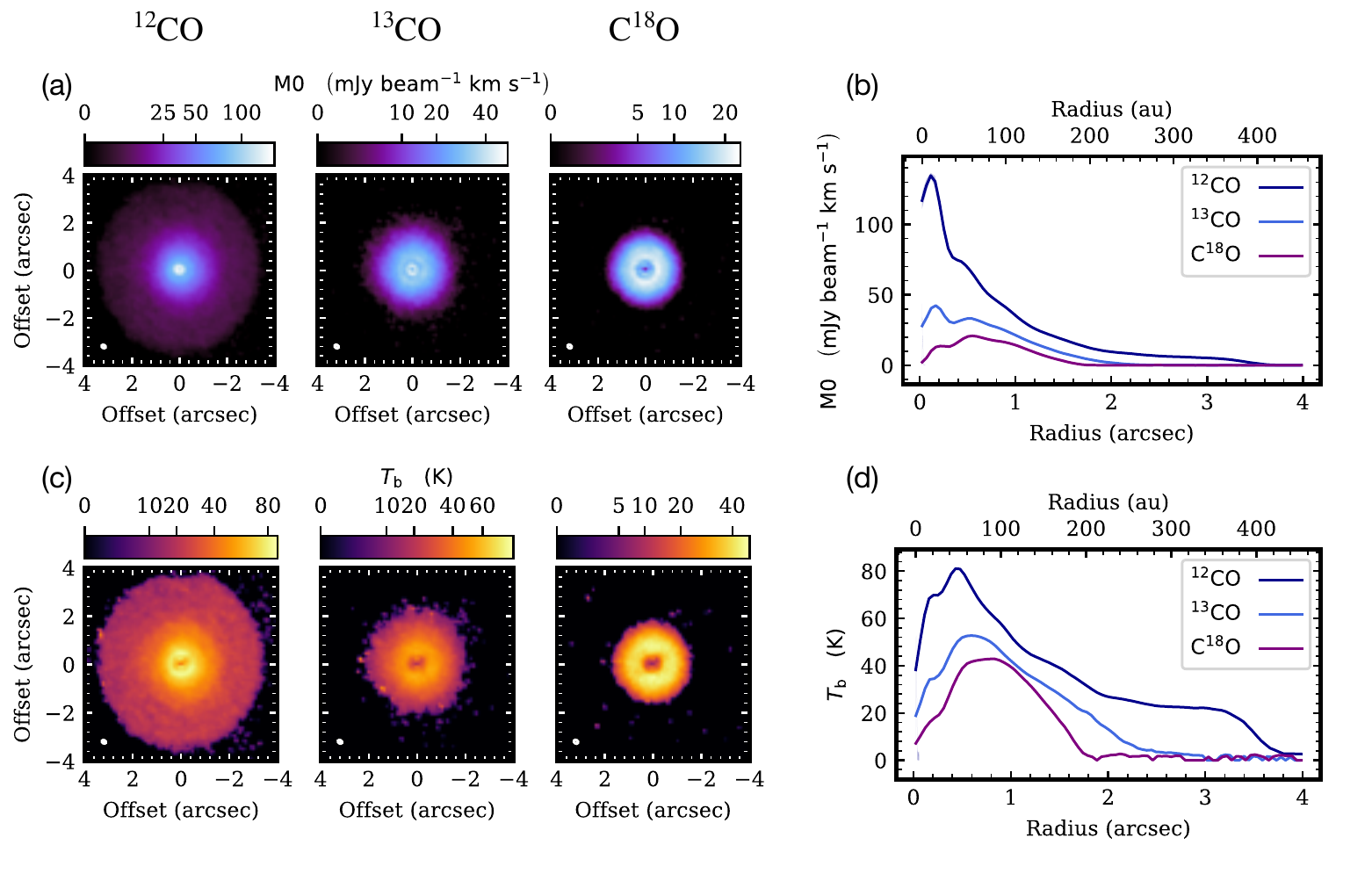}
    \caption{An overview of the gas observation data. Panels (a) and (c) show the Moment 0 maps and the brightness temperature maps of $^{12}$CO (2-1), $^{13}$CO (2-1) and C$^{18}$O (2-1), respectively, while panels (b) and (d) show their radial profiles, respectively. The shaded region in each radial profile shows the azimuthal scatter in each radial bin.}
    \label{fig:ov-gas1}
\end{figure*}

We calculate the Moment 0 (velocity-integrated intensity) map of $^{12}$CO (2-1), $^{13}$CO (2-1) and C$^{18}$O (2-1) with the \texttt{bettermoments}\footnote{\url{https://github.com/richteague/bettermoments}} \citep{2018RNAAS...2c.173T} package, using a clip of 2 $\times$ RMS, shown in Figure~\ref{fig:ov-gas1}a. Their radial profiles are shown in Figure~\ref{fig:ov-gas1}b. We also calculate the brightness temperature maps by measuring the peak values of the line spectra and apply the Planck's law. The maps and their radial profiles are shown in Figure~\ref{fig:ov-gas1}c and d.

\begin{figure*}
    \centering
    \includegraphics[width=0.9\textwidth]{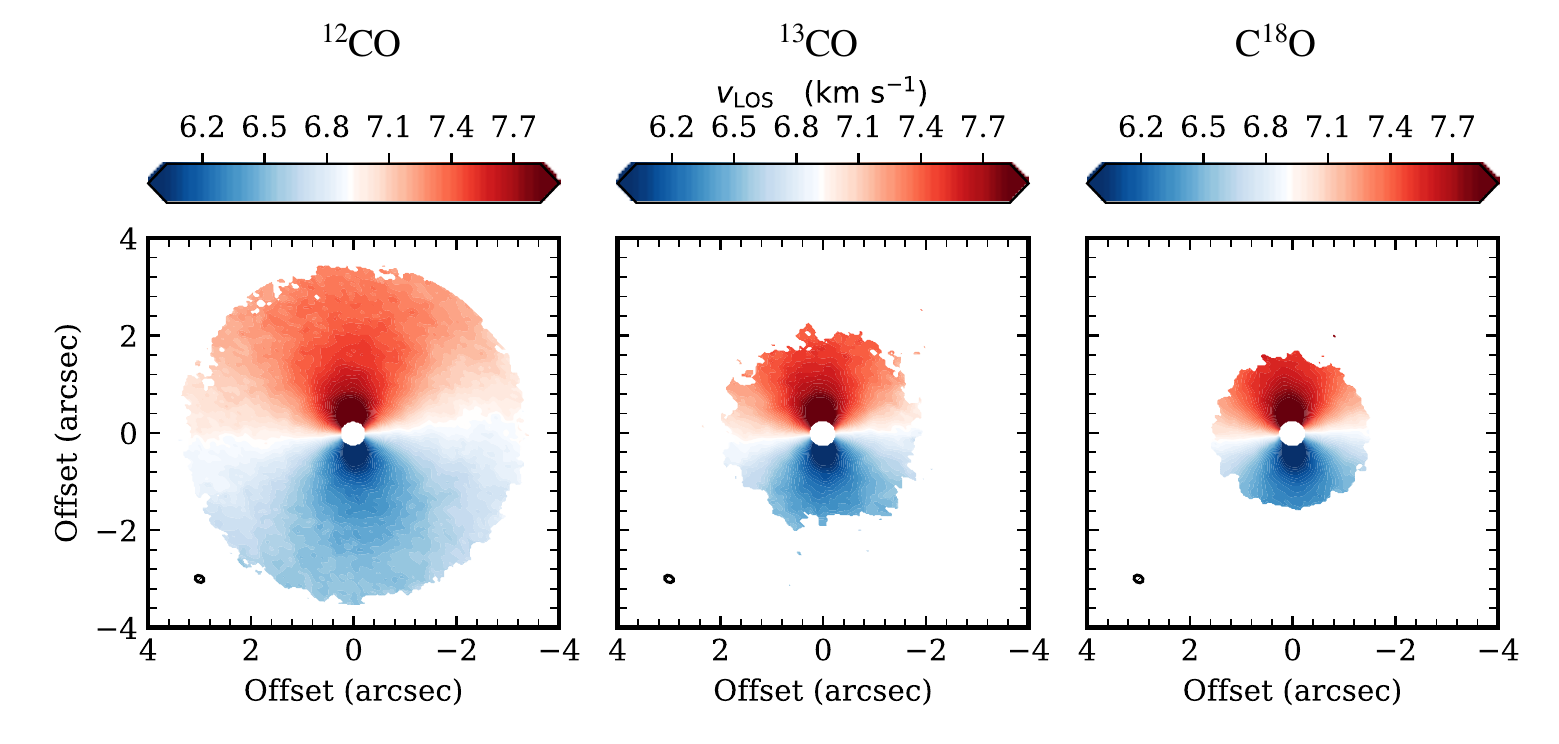}
    \caption{Line-of-sight velocity maps of $^{12}$CO (2-1), $^{13}$CO (2-1) and C$^{18}$O (2-1).}
    \label{fig:ov-gas2}
\end{figure*}

We then calculate the line-of-sight velocity maps using the \texttt{bettermoments} package and employing the \texttt{gaussian} method for all three CO isotopologues, shown in Figure~\ref{fig:ov-gas2}. The \texttt{gaussian} method fits a Gaussian profile to the spectrum at each pixel to measure the line center, $v_{\mathrm{LOS}}$. We have compared different methods for calculating the $v_{\mathrm{LOS}}$ maps and found that the \texttt{gaussian} method provides the lowest statistical uncertainties on the measured $v_{\mathrm{LOS}}$, based on the uncertainties returned from \texttt{bettermoments}. More details on the comparison of different methods can be found in Appendix \ref{apd:1}. 

\section{Measuring the three-dimensional velocity structure}
\label{sec:measuring v}

In this section, we disentangle the 3D velocity structure of the disk from the measured line-of-sight velocity map following \citet{2018ApJ...860L..12T,2018ApJ...868..113T,2019Natur.574..378T}. We extract perturbations in the rotational velocity, and map out large scale gas flows in the radial and vertical direction.

\subsection{Method} \label{M31}
\subsubsection{General source properties} \label{M311}
In order to accurately measure the profiles for three-dimensional velocities, we need geometrical properties, e.g., inclination, positional angle, of the disk to properly deproject the data cubes and split it into annuli of constant radius. We fit the projected line-of-sight velocities, $v_{\mathrm{LOS}}$, with a Keplerian velocity structure to measure the needed geometrical properties.

We use the \texttt{eddy}\footnote{\url{https://github.com/richteague/eddy}} \citep{2019JOSS....4.1220T} package to fit a Keplerian rotation model, i.e., Equation (\ref{eq:4}), without the pressure gradient term, to the velocity maps. As HD 169142 is observed at a near face-on orientation ($i=13\degr$; \citealt{2006AJ....131.2290R}), we assume that the disk is geometrically thin, i.e., setting $z = 0$. This assumption results in an inferred dynamical mass slightly higher than the true stellar mass as we do not account for the vertical dependence of the rotation \citep[e.g.,][]{2012ApJ...757..129R}. If the emission traces a constant $z/r$ across the radius of the disk, then the dynamical mass will be under-estimated by a factor of $\left[1+(z / r)^{2}\right]^{-3 / 2}$. 

We use a Markov chain Monte Carlo (MCMC) method with 256 walkers, 2000 burn-in steps and 10000 steps to estimate the posterior distributions of disk center offset, relative to the image center, $(\Delta x, \Delta y)$, position angle (PA), dynamical mass $M_{\mathrm {star}}$ and systemic velocity $v_{\mathrm{LSR}}$. The best-fit models for each CO isotopologue are described in Table \ref{tab:Kep}. The values shown are the median of the posterior distributions, with statistical uncertainties given by the 16th to 84th percentile range (note, however, that the statistical uncertainties will substantially under-predict the true uncertainties which are rather likely dominated by systematic uncertainties associated with the choice of model, i.e., a Keplerian rotation model). The values we get from the three CO isotopologues are consistent among one another, suggesting that the values we derive are robust.

\begin{deluxetable*}{cccccc}
\tablecaption{Best-fit values of parameters from the Keplerian fitting. The values shown are the median values of the posterior distributions, and the statistical uncertainties are given by the 16th to 84th percentile range. \label{tab:Kep}}
\tablewidth{1pt}
\tablehead{
& $\Delta x$ & $\Delta y$ & PA & $M_{\mathrm {star}}$ & $v_{\mathrm{LSR}}$ \\ 
& (milliarcsec) & (milliarcsec) & ($\degr$) & ($M_{\mathrm {sun}}$) & ($\mathrm{km} \ \mathrm{s}^{-1}$)
}
\startdata
${ }^{12} \mathrm{CO \ (2-1)}$ & $10.5_{-0.3}^{+0.3}$ & $-15.7_{-0.3}^{+0.3}$ & $5.34_{-0.02}^{+0.02}$ & $1.393_{-0.001}^{+0.001}$ & $6.9440_{-0.0002}^{+0.0002}$ \\
${ }^{13} \mathrm{CO \ (2-1)}$ & $18.1_{-0.5}^{+0.5}$ & $-19.6_{-0.5}^{+0.5}$ & $5.34_{-0.03}^{+0.03}$ & $1.418_{-0.002}^{+0.002}$ & $6.9512_{-0.0003}^{+0.0003}$ \\
$\mathrm{C}^{18} \mathrm{O \ (2-1)}$ & $16.6_{-0.4}^{+0.4}$ & $-24.3_{-0.4}^{+0.4}$ & $5.45_{-0.03}^{+0.03}$ & $1.423_{-0.001}^{+0.001}$ & $6.9487_{-0.0003}^{+0.0003}$ \\
\enddata
\end{deluxetable*}

\subsubsection{Measuring rotational, radial, and vertical velocities}

To decompose the projected line-of-sight velocities, $v_{\mathrm{LOS}}$, into their cardinal components, we follow the method outlined in \citet{2018ApJ...860L..12T,2018ApJ...868..113T,2019Natur.574..378T}. In essence, we are able to disentangle the three velocity components due to their different dependence on the disk azimuthal angle, $\phi$. We first use the \texttt{GoFish} package to measure the azimuthally averaged profiles for the rotational velocity, $v_{\mathrm{rot}}$, and the radial velocity, $v_{\mathrm{rad}}$. This first splits the data cube into annuli each 25 mas wide, roughly a quarter of the beam, based on the geometrical properties shown in Table \ref{tab:Kep}. In each annulus, spatially independent spectra are shifted to a common velocity given a $\{v_{\rm rot},\, v_{\rm rad}\}$ tuple. The aligned spectra are compared to a Gaussian Process (GP) model to extract a goodness of fit. The best-fit values for $v_{\mathrm{rot}}$ and $v_{\mathrm{rad}}$ should minimize the variance in residuals between the average spectrum and the GP model. As the vertical velocity, $v_{\mathrm{z}}$, has no dependence on $\phi$, it will not contribute to this variance. We use MCMC with 32 walkers, 500 burn-in steps and 500 steps on each annulus to explore the posterior distributions of model parameters of $v_{\mathrm{rot}}$ and $v_{\mathrm{rad}}$ to estimate the best-fit values. To better characterize the range of velocity profiles consistent with the data, we run multiple instances of the above fitting procedure, each time selecting a random set of independent pixels, and taking the mean and standard deviation of the ensemble of samples as the best-fit value and associated uncertainty (as in \citealt{2019A&A...625A.118K}). The final profile is a combination of 25 unique pixels selections. Appendix \ref{apd:21} describes in more detail how the number of independent samples was chosen.

We then estimate the profiles of vertical velocity, $v_{\mathrm{z}}$, in a different way to \citet{2019Natur.574..378T}. In \citet{2019Natur.574..378T}, the line center of the velocity-corrected and stacked line profiles is determined by fitting a Gaussian profile. In this work, following Equation (\ref{eq:vlos}), subtracting the inferred $v_{\mathrm{rot}}$ and $v_{\mathrm{rad}}$ profiles, together with the $v_{\mathrm{LSR}}$, from the $v_{\mathrm{LOS}}$ map will leave a residual map of $v_{\mathrm{z}}$ values. We then make an azimuthally averaged radial profile of this residual map to measure the radial profile of $v_{\mathrm{z}}$. This approach is preferable to the approach used in \citet{2019Natur.574..378T} as we circumvent any issues associated with asymmetries introduced in the line profiles from poor constraints on $v_{\mathrm{rot}}$ or $v_{\mathrm{rad}}$ (see the discussion on asymmetric line profiles in the disk in \citealt{2020ApJ...899..157T}). It was verified for the case of HD~169142 that, as there was no asymmetry in the line profiles, both methods returned comparable profiles for $v_{\rm z}$.

\subsection{Results} \label{M32}

\begin{figure*}
    \centering
    \includegraphics[width=1.0\textwidth]{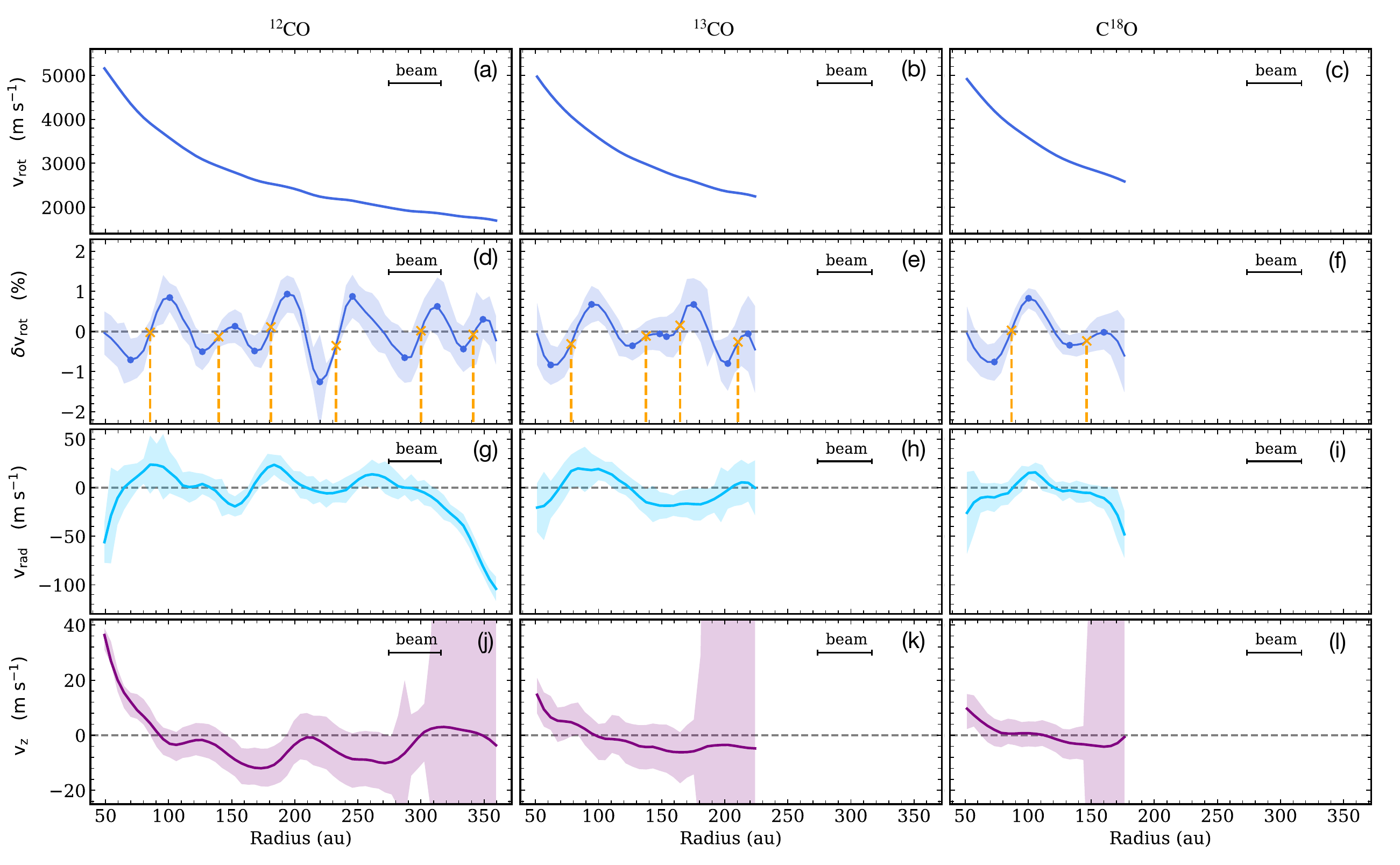}
    \caption{Velocity profiles of ${ }^{12} \mathrm{CO \ (2-1)}$, ${ }^{13} \mathrm{CO \ (2-1)}$, and $\mathrm{C}^{18} \mathrm{O \ (2-1)}$, are shown in left, middle, and right columns, respectively. The first, third and forth rows show the profiles for the rotational velocity, $v_{\mathrm{rot}}$, radial velocity, $v_{\mathrm{rad}}$, and vertical velocity, $v_{\mathrm{z}}$, respectively. The second row shows the perturbations in $v_{\mathrm{rot}}$ as a percentage of the background rotational velocity, $\delta v_{\mathrm{rot}}$. The orange crosses and dotted lines in the $\delta v_{\mathrm{rot}}$ profiles show the indicated local pressure minima. The shaded areas in each profile show the uncertainties.}
    \label{fig:2}
\end{figure*}

The resulting velocity profiles are shown in Figure \ref{fig:2}. The shaded areas show the uncertainties of each profile. The first, third and fourth rows show the profiles for the rotational velocity, $v_{\mathrm{rot}}$, radial velocity, $v_{\mathrm{rad}}$, and vertical velocity, $v_{\mathrm{z}}$, respectively. The second row shows the perturbations in $v_{\mathrm{rot}}$ as a percentage of the background rotational velocity, $\delta v_{\mathrm{rot}}$. Note that all the velocity profiles start at around 50~au. This is because these sort of kinematic techniques are only possible at radii more than twice the beam major axis away from the center, which was tested with forward models in \citet{2018ApJ...860L..12T}. For current observations, this is equivalent to between 40 and 45~au.

To identify the perturbations, $\delta v_{\mathrm{rot}}$, we implement a Butterworth low-pass filter, with a cut-off of 16 samples ($\sim$83 au) and an order of 12, to extract a smooth background model of $v_{\mathrm{rot}}$. \citet{2019Natur.574..378T} shows that assuming different background models will lead to subtle differences in the low-frequency (i.e., large spatial scale) perturbations, although high-frequency perturbations (small spatial scale features) persist. The filter technique used here will minimize such risk as it distinguishes `background' and `perturbation' based on their spatial frequency, rather than the assumption that the background can be well described by an analytical model. Appendix~\ref{apd:22} discussed this in more detail.

As shown in the second row of Figure \ref{fig:2}, this approach reveals perturbations in $v_{\mathrm{rot}}$ on scales of tens of au with magnitudes of around 1 - 2\% of the background rotation. Given that the process of estimating the $v_{\mathrm{rot}}$ profiles includes pixel shifting which decorrelates the noise in the emission profiles \citep{2016ApJ...832..204Y}, we are sensitive to variations on scales smaller than the beam size. While the perturbations in the $^{12} \mathrm{CO}$ seems to have a constant spatial frequency similar to that of the beam FWHM, this is not the case for $^{13} \mathrm{CO}$ and $\mathrm{C}^{18} \mathrm{O}$ suggesting that it is just a coincidence.

As discussed in \citet{2015MNRAS.448..994K}, positive gradients in $\delta v_{\mathrm{rot}}$ indicate local minima in gas pressure (gaps), which might further indicate interesting processes (e.g., embedded planets) occurring around those locations. Following \citet{2020MNRAS.495..173R}, we assume the local minimum in the gas pressure are at the middle point between the local minima and maxima in $\delta v_{\mathrm{rot}}$. We label such locations with orange crosses and dotted lines in the $\delta v_{\mathrm{rot}}$ profiles. \citet{2018ApJ...860L..12T} is the first to demonstrate the technique to predict the gas pressure profile from the measured rotation curves. Their application to CO isotopologue emission from HD 163296 disk revealed predicted local minima of the gas pressure that are aligned with the gaps in the continuum. A more comprehensive study of gas properties through $v_{\mathrm{rot}}$ profiles will be described in Section \ref{sec:model}.

\begin{figure*}
    \centering
    \includegraphics[width=0.8\textwidth]{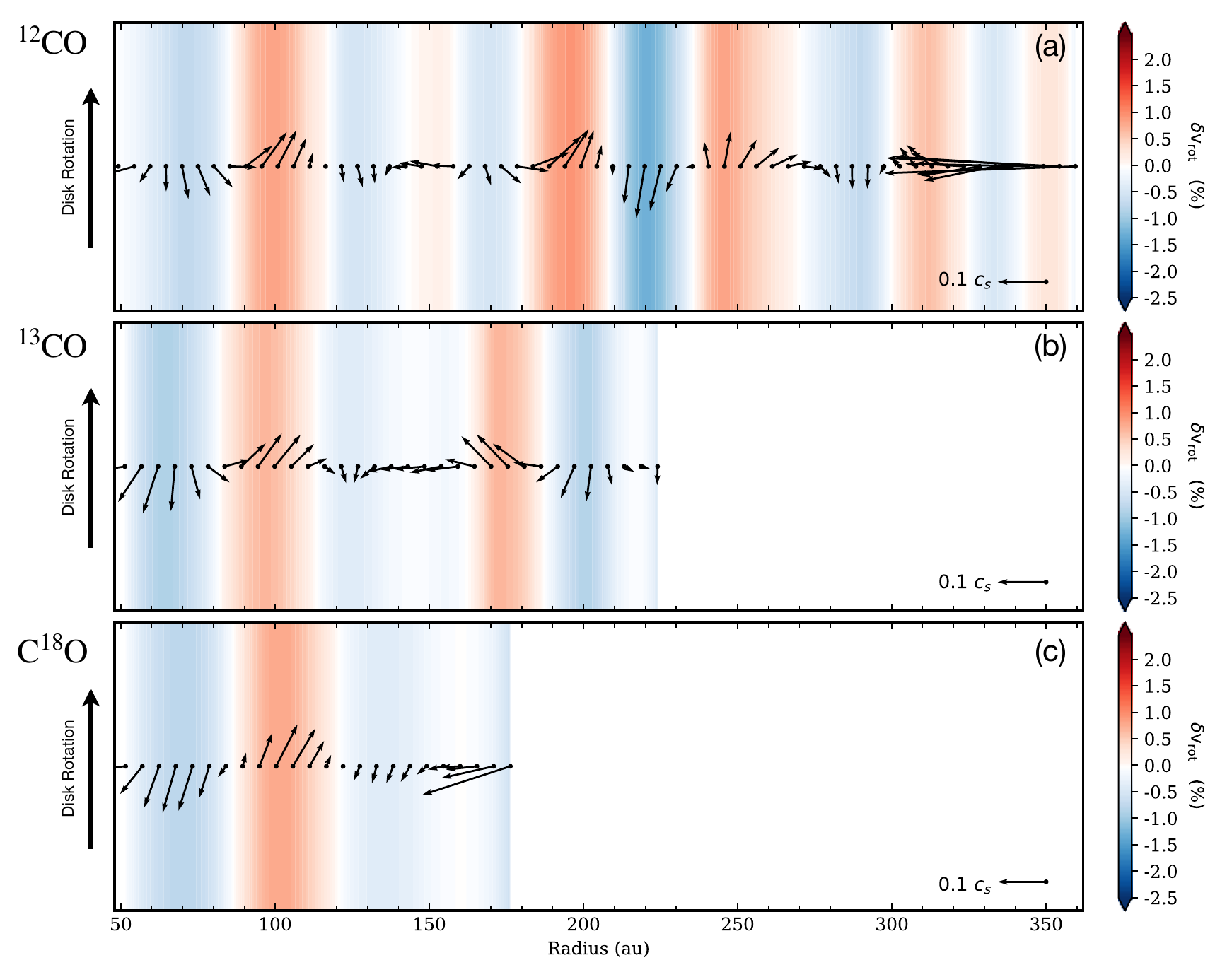}
    \caption{Panels (a), (b), and (c) represent gas flows in the $(r, \phi)$ plane of the three CO isotopologues. The colour of the background shows the magnitude of the perturbations in the rotational velocity. A vector at the bottom right in each panel shows 0.1 $c_{\mathrm{s}}$.}
    \label{fig:31}
\end{figure*}

\begin{figure*}
    \centering
    \includegraphics[width=0.9\textwidth]{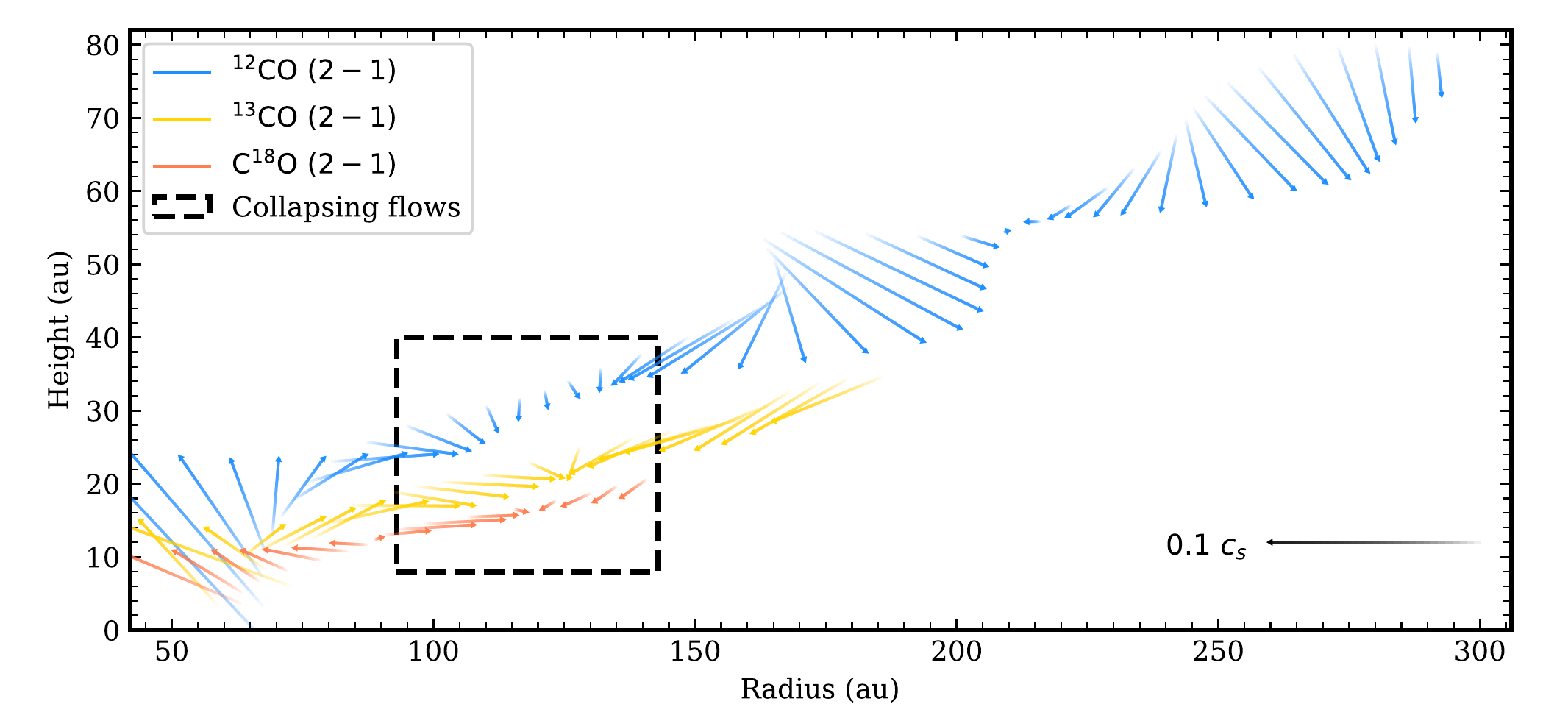}
    \caption{Gas flows in the $(r, z)$ plane of the three CO isotopologues. The black dotted box indicates the location of the coherent collapsing gas flow at around 125 au. A vector at the bottom right shows 0.1 $c_{\mathrm{s}}$.}
    \label{fig:32}
\end{figure*}

Inferred velocity structures, rescaled as a function of the local sound speed, $c_{\mathrm{s}}$, are shown in Figure \ref{fig:31} and \ref{fig:32}. Given that the three emission lines we use are optically thick, we adopt the observed brightness temperature, shown in Figure \ref{fig:ov-gas1}c and d, as the local gas temperature for calculating the local sound speed. Note, however, that the temperatures are very low (i.e., $\leq$ 20~K) in the outer disk (i.e., $\geq$ 200~au) for all three isotopologues and the very inner disk for $\mathrm{C}^{18} \mathrm{O}$. These probably suggest that the lines may not be optically thick in these regions as CO is expected to only be in the gas phase above 20~K. Thus we use 20~K to calculate the local sound speed when it is indicated to be $\leq$ 20~K. Panels (a), (b), and (c) of Figure \ref{fig:31} show the gas flows in the $(r, \phi)$ plane traced by the three CO isotopologues. Arrows pointing upwards represent \emph{super}-Keplerian motion, while arrows pointing downwards show \emph{sub}-Keplerian motion. The colour of the background shows the magnitude of the perturbations of the rotational velocity relative to the background rotation. We can see that the perturbations of the rotation velocities of the three CO isotopologues are roughly consistent with each other. Note that at around 200~au, the perturbations of $^{12} \mathrm{CO}$ and $^{13} \mathrm{CO}$ are misaligned with each other. This is potentially due to the inability of estimating of the emission height, as this disk is observed almost at a face-on orientation, resulting in an incorrect calculation of radial distance, $r$. 

Figure \ref{fig:32} shows the gas flows in the $(r, z)$ plane traced by all three CO isotopologues. Given that we are not able to extract the emission height due to the face-on geometry of the system, we assume emission heights based on the disk models shown in \citet{2017A&A...600A..72F}: for $^{12} \mathrm{CO \ (2-1)}$, $z/r = 0.26$; for $^{13} \mathrm{CO \ (2-1)}$, $z/r = 0.18$; for $\mathrm{C}^{18} \mathrm{O \ (2-1)}$, $z/r = 0.14$. These values of emission heights are broadly consistent with those observed in other sources, such as for the disk around IM Lup \citep{2018A&A...609A..47P,2021arXiv210906217L}. Note that the values of the emission height used here do not affect the inferred velocity structure, they are merely used for plotting the figure. In Fig.~\ref{fig:32}, we can clearly see collapsing gas flows around 125 au (indicated by a black dotted box), and tentatively 210 au, reminiscent of meridional flows \citep{2014ApJ...782...65S}. Around these locations, gas flows toward a common center, and downward towards the disk mid-plane. Both locations are well beyond the outer edge of the continuum emission, and so unaffected by any continuum subtraction. Strikingly, we see coherent flows traced by all three CO isotopologues at 125 au -- the first time coherent flows have been traced across a range of heights in the disk. The interpretation of these flows will be discussed in Section \ref{sec:discussion}.

\section{Inferring the perturbed\\physical structure} \label{sec:model}
To account for the deviations observed in the rotational velocity (Figure \ref{fig:2}, second row), we require perturbations in the gas number density, temperature or emission height, e.g., Eqn.~\ref{eq:4}. In reality, this is likely a combination of all three properties, requiring full thermo-chemical modeling \citep[e.g.,][]{2020A&A...642A.165R}. As a first step, we infer the perturbations in each component assuming that only one component is perturbed at a time, as in \citet{2018ApJ...868..113T}.

\subsection{Method} \label{M41}
For a geometrically thick disk, the rotational velocity $v_{\mathrm{rot}}$ is given by Equation (\ref{eq:4}). To account for the deviations observed in $v_{\mathrm{rot}}$, we need perturbations in either the CO emission height, $z_{\mathrm{CO}}$, or the background gas density (here the \emph{total} gas density, not just CO gas) and temperature at the heights traced by the CO isotopologue emission, $n(\mathrm{H_2})_{z=z_{\mathrm{CO}}}$ and $T_{z=z_{\mathrm{CO}}}$, respectively.

For temperature, $T_{z=z_{\mathrm{CO}}}$, we assume that all three isotopologues are optically thick, such that we can fix the temperature to the observed brightness temperature shown in Figure \ref{fig:ov-gas1}c and d. Note that even in the case the lines are not optically thick, their intensities (or brightnesses) should still be linearly dependent on the temperature, and thus will not change the magnitude of the perturbations relative to their background value. For a more robust numerical differentiation, we require a smoothly varying model of $T_{z=z_{\mathrm{CO}}}$, otherwise small noise spikes will manifest as spurious features in the pressure gradient and thus the velocity profile. We use a Savitzky-Golay filter with a window of 9 data points ($\sim$ 47 au) and polynomial order of 5 to remove the high frequency noise in the observed $T_\mathrm{B}$ profiles, before using cubic interpolation to extract an interpolated model. Note that the process will not smooth out real perturbations originated in the observed $T_\mathrm{B}$ profiles. It only removes high frequency signal which has a much smaller scale than the beam size, and are thus noise.

For $n(\mathrm{H_2})_{z=z_{\mathrm{CO}}}$ and $z_{\mathrm{CO}}$, we construct models of the rotational velocity by considering smooth `background' models each of which are perturbed with a sum of multiple Gaussian components. We refer to this sum of Gaussian components as the `perturbation'. \citep[following][for example]{2018ApJ...868..113T}.

For the smooth background of emission height, $z_{\mathrm{CO}}$, we model it as having a constant $z/r$ for each isotopologue. We adopt the values described in Section \ref{M32}: $^{12} \mathrm{CO \ (2-1)}$, $z_{\mathrm{bkg}}/r = 0.26$; $^{13} \mathrm{CO \ (2-1)}$, $z_{\mathrm{bkg}}/r = 0.18$; $\mathrm{C}^{18} \mathrm{O \ (2-1)}$, $z_{\mathrm{bkg}}/r = 0.14$. 

For the smooth background of density profile, $n(\mathrm{H_2})_{z=z_{\mathrm{CO}}}$, we opt to use a tapered-power law,
\begin{equation} \label{eq:5}
n_{\mathrm{bkg}}(r)= n_{\mathrm{0}} \ \left(\frac{r}{100~\mathrm{au}}\right)^{q_{\mathrm{n1}}} \ \exp \left[-\left(\frac{r}{r_{{\rm exp}\, n}}\right)^{q_{\mathrm{n2}}}\right],
\end{equation}
\noindent which was found through testing various analytical forms to provide the best trade off between flexibility and a minimal number of free parameters: $q_{\mathrm{n1}}$, $q_{\mathrm{n2}}$, $r_{{\rm exp}\, n}$, and $n_{\mathrm{0}}$. Given that $n_{\mathrm{0}}$ in Equation (\ref{eq:5}) is cancelled out during the calculation of the modeled $v_{\mathrm {rot}}$, we are unable to constrain the absolute scaling of $n_{\mathrm{bkg}}(r)$. Hence, we use normalized values for $n(\mathrm{H_2})_{z=z_{\mathrm{CO}}}$ profiles.

We model the perturbation vector, $\delta (r)$, as the summation of between six and ten Gaussians,
\begin{equation}
\delta (r) = 1.0 + \sum_{i=1}^{k} \left\{A_{i} \exp \left[-0.5{\left(\frac{r-r_{0i}}{d r_{i}}\right)}^{2}\right]\right\}.
\end{equation}
\noindent The numbers of Gaussian components used, $k$, depend on the numbers of the perturbations seen in the $\delta v_{\mathrm{rot}}$ profiles. For $^{12} \mathrm{CO \ (2-1)}$, $k = 11$; for $^{13} \mathrm{CO \ (2-1)}$, $k = 7$; for $\mathrm{C}^{18} \mathrm{O \ (2-1)}$, $k = 6$. We multiply the perturbations, $\delta (r)$, with the background models to construct the fully perturbed models. We are then able to calculate the perturbed $v_{\mathrm{rot}}$ profile from the perturbed model of $n(\mathrm{H_2})_{z=z_{\mathrm{CO}}}$ or $z_{\mathrm{CO}}$ through Equation (\ref{eq:4}). For the calculation of $v_{\mathrm{rot}}$, we calculate the pressure profile on a radial grid that is sampled at a rate 10 times higher than the observed values, such that we can avoid any finite-differencing issues in the calculation of the gradient.

To obtain the best-fit values of the free parameters in the perturbed models of $n(\mathrm{H_2})_{z=z_{\mathrm{CO}}}$ or $z_{\mathrm{CO}}$, we first manually find reasonable initial values for these parameters. Then we use the \texttt{dynesty}\footnote{\url{https://github.com/joshspeagle/dynesty}} package \citep{2020MNRAS.493.3132S}, which implements dynamic nested sampling \citep{higson_dynamic_2019} to explore the posterior distributions of the parameters around their initial values consistent with the observed $v_{\mathrm{rot}}$ profiles. We use dynamic nested sampling instead of MCMC methods for the following reasons. Firstly, nested sampling is more robust than MCMC methods when sampling complex, multi-modal distributions \citep[e.g.,][]{2014MNRAS.437.3918A, 2020MNRAS.493.3132S}, such as the parameter space here. Secondly, dynamic nested sampling can be much more efficient than the static nested sampling given that it can dynamically allocate the number of live points (analogous to walkers in an MCMC approach) during a run. This enables us to obtain the highest accuracy per unit running time, particularly important when the total computational costs are high due to, e.g., high dimensionality of the parameter space (to fully modeled the perturbed density, we need around 22 to 37 free parameters, leading to prohibitive computational costs). 

\begin{figure*} 
    \centering
    \includegraphics[width=1.0\textwidth]{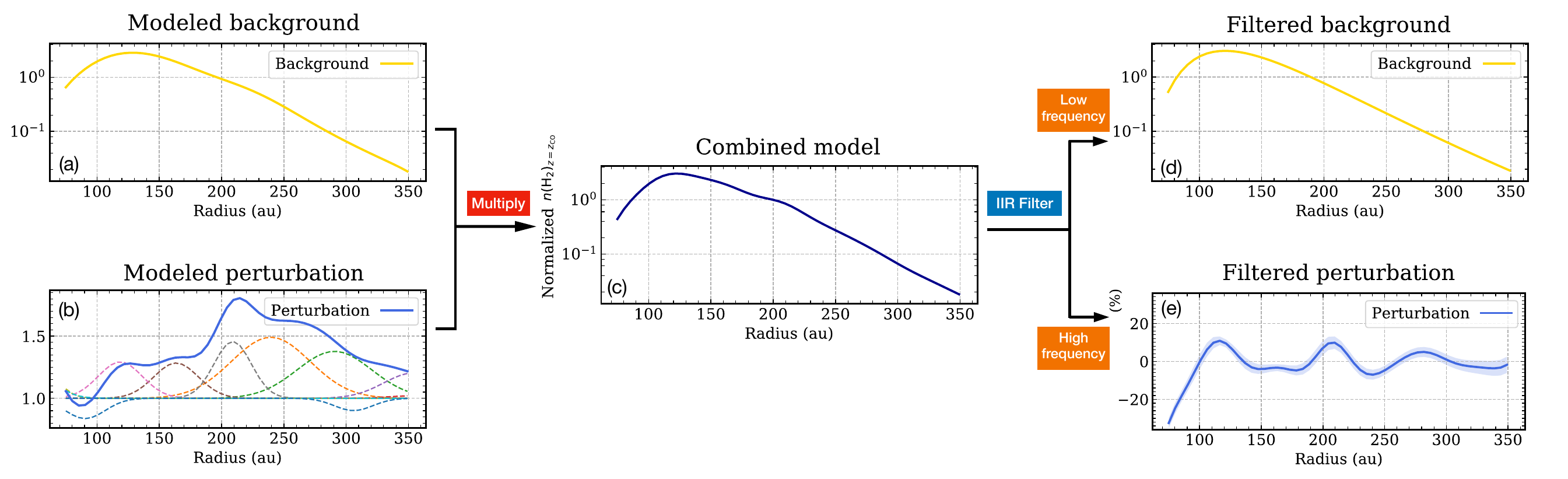}
    \caption{A diagram shows the workflow described in Section \ref{M41}. Panel (a) shows the normalized modeled background of $n(\mathrm{H_2})_{z=z_{\mathrm{CO}}}$. Panel (b) shows the modeled perturbation: the dotted lines show the curve of each Gaussian perturbation and the blue solid line shows the sum of these curves. By multiplying the modeled perturbation to the background, we get the modeled perturbed $n(\mathrm{H_2})_{z=z_{\mathrm{CO}}}$. We then use the dynamic nested sampling to find the best-fit values for the perturbed $n(\mathrm{H_2})_{z=z_{\mathrm{CO}}}$, one example is shown in panel (c). Then we use the above mentioned filtering technique to separate the background and perturbation profiles, shown in panel (d) and (e), respectively.}
    \label{fig:pb-process}
\end{figure*}

Although we model our gas density, emission height as the product of a smooth background and a perturbation term, these two components are not necessarily the same two components extracted when using a filtering technique as was used to identify perturbations in the $v_{\mathrm{rot}}$ profile. Using the same filtering technique on our model output, we are then able to split our model profiles into analogous background and perturbation vectors, such that we are only sensitive to the high-frequency perturbations. Fig.~\ref{fig:pb-process} demonstrates this idea. Panel (a) shows the normalized modeled background of $n(\mathrm{H_2})_{z=z_{\mathrm{CO}}}$. Panel (b) shows the modeled perturbations: the dotted lines show the curve of each Gaussian perturbation and the blue solid line shows the sum of these curves. By multiplying the modeled perturbation to the background, we get the modeled perturbed $n(\mathrm{H_2})_{z=z_{\mathrm{CO}}}$, shown in panel (c). Then we use the above mentioned filtering technique to separate the background and perturbation profiles, shown in panels (d) and (e), respectively.

We follow this approach for the $n(\mathrm{H_2})_{z=z_{\mathrm{CO}}}$, $z_{\mathrm{CO}}$ separately, and for each CO isotopolgoue independently. The coupling between $n(\mathrm{H_2})_{z=z_{\mathrm{CO}}}$, $z_{\mathrm{CO}}$, and $T_{z=z_{\mathrm{CO}}}$ will be discussed in Section \ref{sec:discussion}.

\subsection{Results} \label{M42}
\begin{figure*} 
    \centering
    \includegraphics[width=1.0\textwidth]{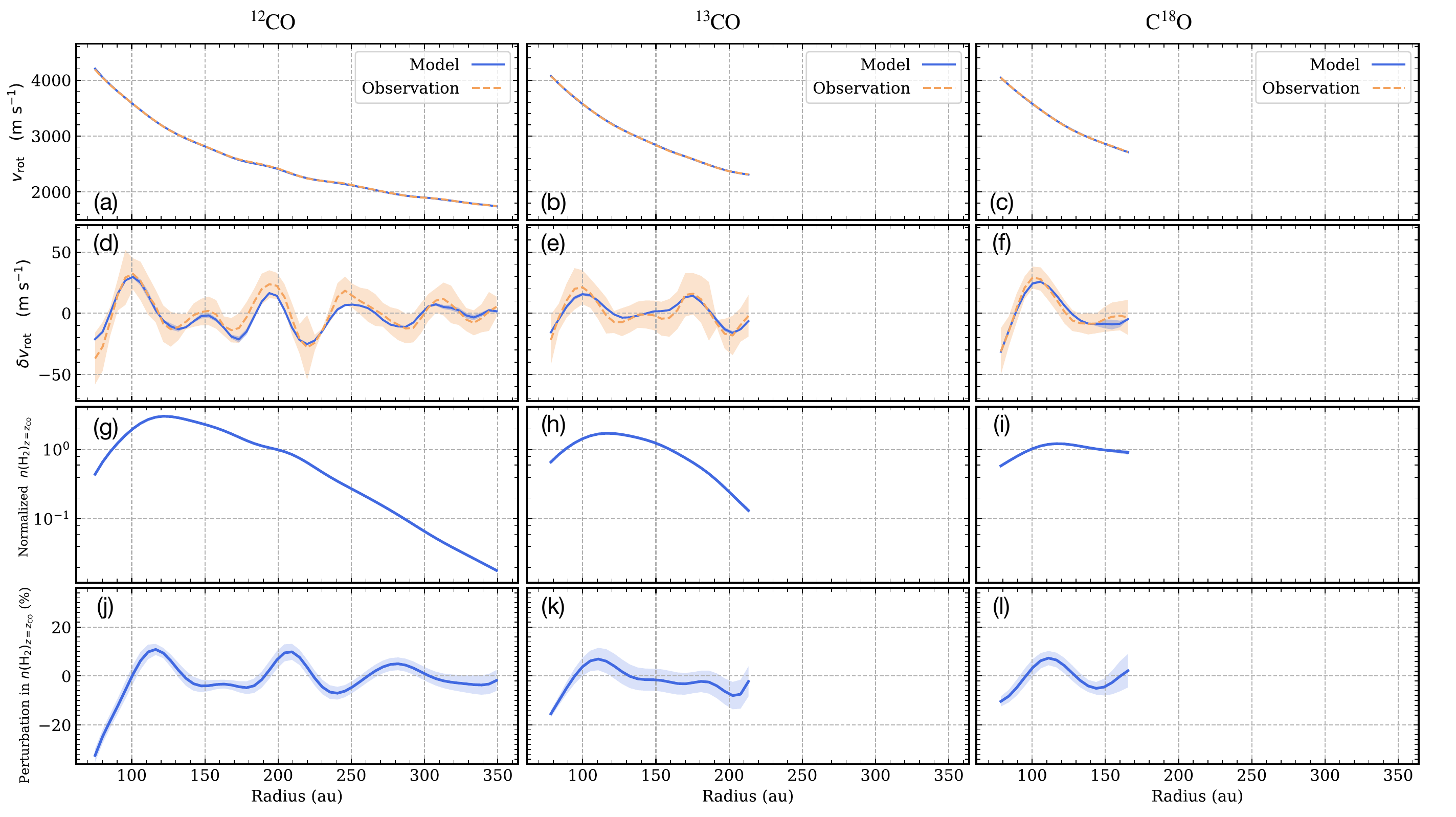}
    \caption{Estimation of the perturbed profiles of gas number density traced by $^{12} \mathrm{CO \ (2-1)}$, $^{13} \mathrm{CO \ (2-1)}$, and $\mathrm{C}^{18} \mathrm{O \ (2-1)}$, are shown in left, middle, and right columns, respectively. Panels (a), (b), and (c) show comparisons between observed $v_{\mathrm{rot}}$ (orange), and modeled perturbed $v_{\mathrm{rot}}$ (blue). Panels (d), (e), and (f) show comparisons between the deviations in observed (orange) and modeled (blue) perturbed $v_{\mathrm{rot}}$, which are extracted using the same filtering technique mentioned in Section \ref{M32}. The orange shaded area shows the error of the observed $v_{\mathrm{rot}}$. Panels (g), (h) and (i) show the estimated normalized perturbed $n(\mathrm{H_2})_{z=z_{\mathrm{CO}}}$. Panels (j), (k), and (l) show the extracted perturbations of the perturbed $n(\mathrm{H_2})_{z=z_{\mathrm{CO}}}$ using the same filtering technique mentioned in Section \ref{M32}.}
    \label{fig:pt-pf-n}
\end{figure*}

The inferred perturbed profiles of the gas number density, $n(\mathrm{H_2})_{z=z_{\mathrm{CO}}}$, are shown in Figure \ref{fig:pt-pf-n} for each CO isotopologue. The estimated perturbed $n(\mathrm{H_2})_{z=z_{\mathrm{CO}}}$ shown here are the weighted mean values of all samples from the posterior distributions, where the weight is the importance weight of each sample in the dynamic nested sampling. The errors, indicated by the blue shaded area around them, are similarly calculated from the weighted standard deviation of all samples. Panels (a), (b), and (c) show comparisons between observed $v_{\mathrm{rot}}$ (orange), and modeled perturbed $v_{\mathrm{rot}}$ (blue). Panels (d), (e), and (f) show comparisons between the deviations from the observations (orange, with associated uncertainties) and from using the modeled $v_{\mathrm{rot}}$ (blue). Panels (g), (h) and (i) show the normalized $n(\mathrm{H_2})_{z=z_{\mathrm{CO}}}$ profile that yields a consistent $v_{\mathrm{rot}}$ profile, while panels (j), (k), and (l) show the filter-extracted perturbations in $n(\mathrm{H_2})_{z=z_{\mathrm{CO}}}$.

This modeling suggests a significant depletion in gas number density inwards of 100~au traced by each of the three CO isotopologues (Fig.~\ref{fig:pt-pf-n}g, h, i). Such a density depletion may be the edge of the dust and gas gaps identified to span between 29 and 55 au \citep{2017A&A...600A..72F, 2018MNRAS.474.5105B}. In panels (j), (k) and (l), we find that the locations of the peaks and gaps in the perturbations are roughly consistent among the three CO isotopologues. Particularly, we see a clear coherent peak around 110 to 120~au in all three isotopologues.

\begin{figure*} 
    \centering
    \includegraphics[width=1.0\textwidth]{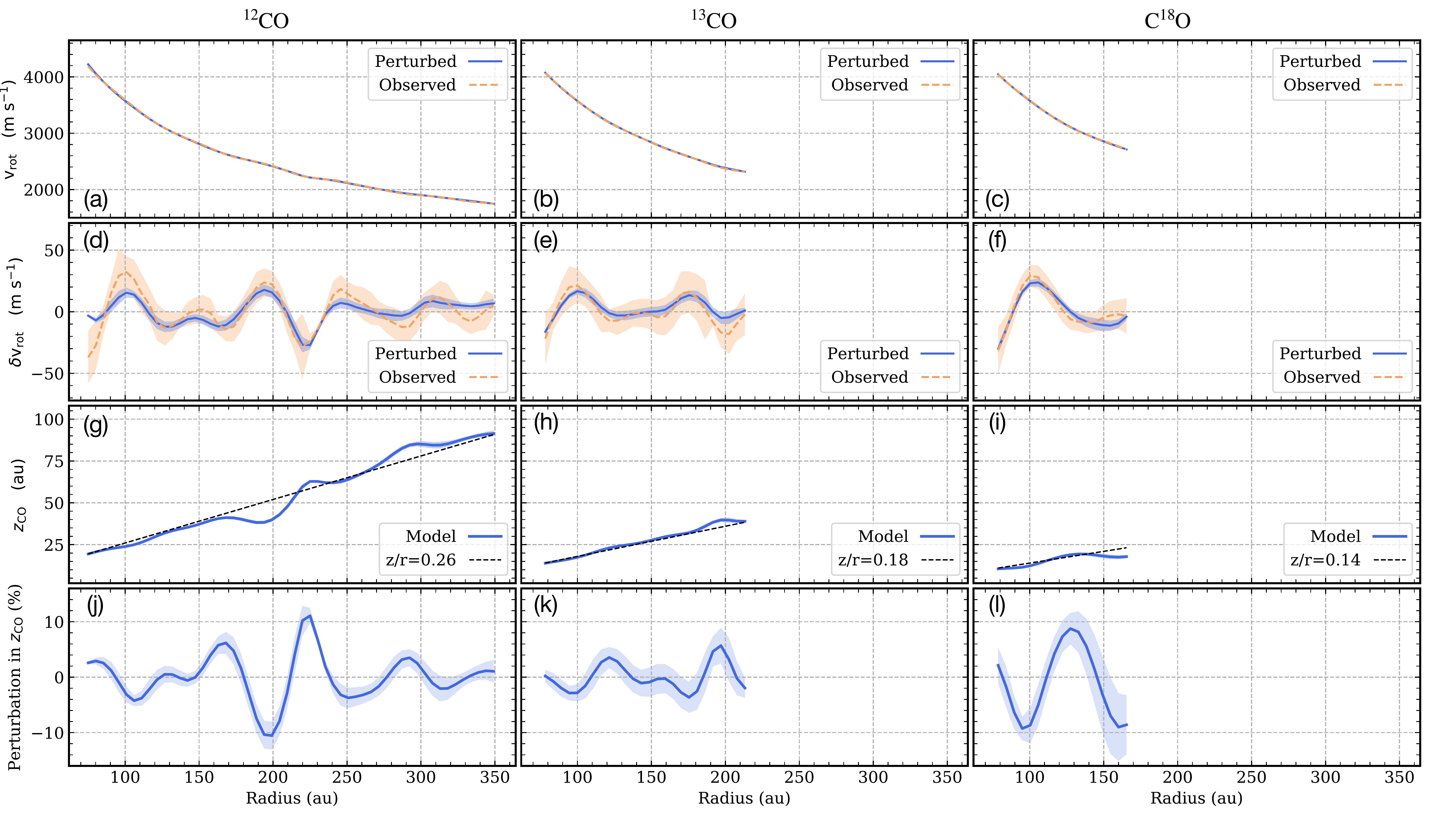}
    \caption{Estimation of the perturbed profiles of emission height traced by $^{12} \mathrm{CO \ (2-1)}$, ${13} \mathrm{CO \ (2-1)}$, and $\mathrm{C}^{18} \mathrm{O \ (2-1)}$, are shown in left, middle, and right columns, respectively. Panels (a), (b), and (c) show comparisons between observed $v_{\mathrm{rot}}$ (orange), and modeled perturbed $v_{\mathrm{rot}}$ (blue). Panels (d), (e), and (f) show comparisons between the deviations in observed (orange) and modeled (blue) perturbed $v_{\mathrm{rot}}$, which are extracted using the same filtering technique mentioned in Section \ref{M32}. The orange shaded area shows the error of the observed $v_{\mathrm{rot}}$. Panels (g), (h) and (i) show the estimated perturbed (blue) $z_{\mathrm{CO}}$. Panels (j), (k), and (l) show the extracted perturbations of the perturbed $z_{\mathrm{CO}}$ using the same filtering technique mentioned in Section \ref{M32}.}
    \label{fig:pt-pf-z}
\end{figure*}

Estimation of the perturbed profiles of emission height, $z_{\mathrm{CO}}$, are shown in Figure \ref{fig:pt-pf-z}, following the same structure of Figure \ref{fig:pt-pf-n}. The locations of the peaks and gaps in the perturbations are roughly consistent among the three CO isotopologues, e.g., we see a coherent gap around 95 to 105~au in all three isotopologues. Comparing to the perturbations in $n(\mathrm{H_2})_{z=z_{\mathrm{CO}}}$, the perturbations in $z_{\mathrm{CO}}$ are roughly at the same location but in a opposite direction, e.g., from $^{12} \mathrm{CO}$, there is a negative perturbation in $z_{\mathrm{CO}}$, but a positive perturbation in $n(\mathrm{H_2})_{z=z_{\mathrm{CO}}}$ at around 200 to 210~au.

\begin{figure*} 
    \centering
    \includegraphics[width=1.0\textwidth]{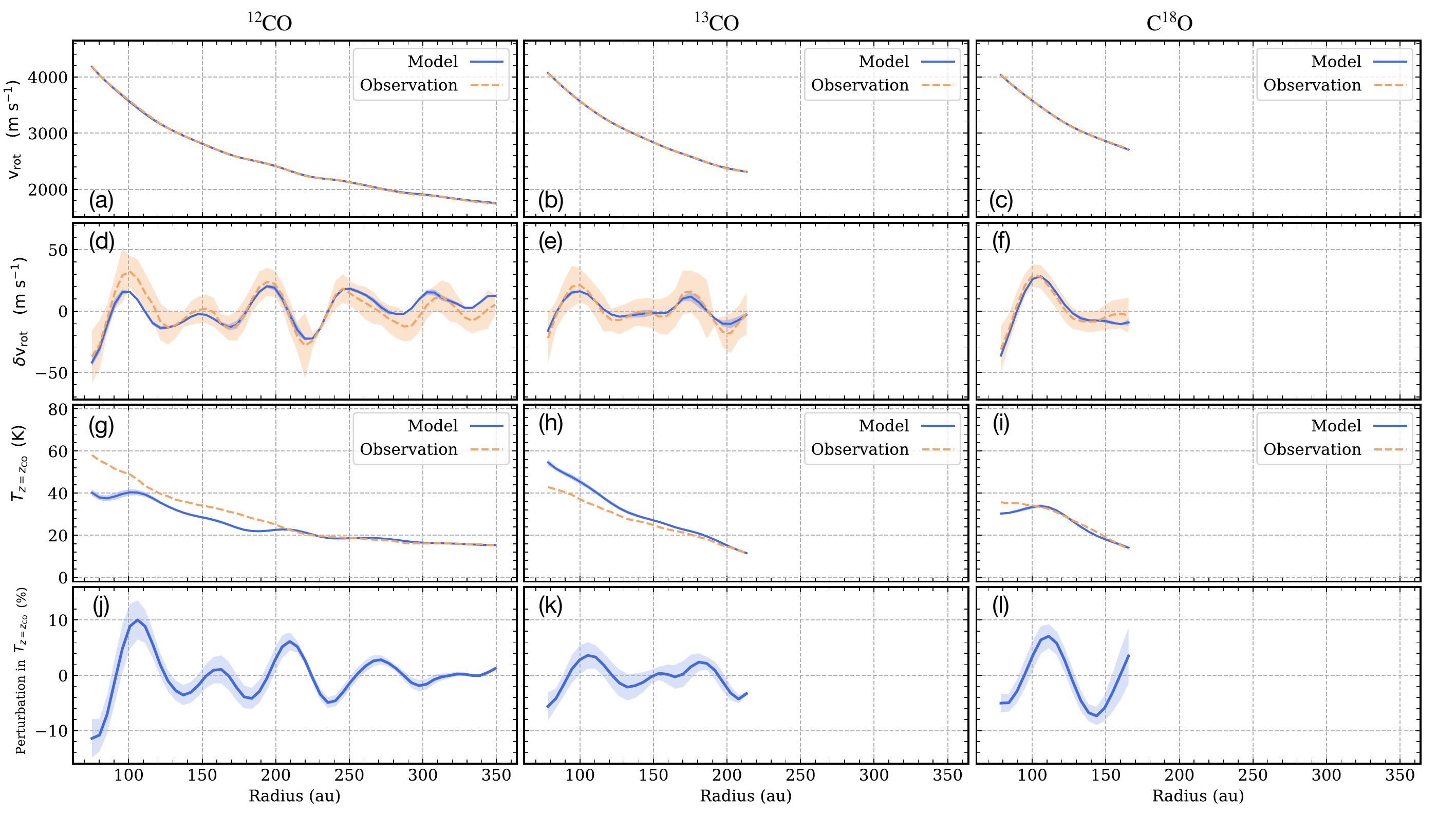}
    \caption{Estimation of the perturbed profiles of local temperature traced by $^{12} \mathrm{CO \ (2-1)}$, $^{13} \mathrm{CO \ (2-1)}$, and $\mathrm{C}^{18} \mathrm{O \ (2-1)}$, are shown in left, middle, and right columns, respectively. Panels (a), (b), and (c) show comparisons between observed $v_{\mathrm{rot}}$ (orange), and modeled perturbed $v_{\mathrm{rot}}$ (blue). Panels (d), (e), and (f) show comparisons between the deviations in observed (orange) and modeled (blue) perturbed $v_{\mathrm{rot}}$, which are extracted using the same filtering technique mentioned in Section \ref{M32}. The orange shaded area shows the error of the observed $v_{\mathrm{rot}}$. Panels (g), (h) and (i) show the comparison between the estimated perturbed (blue) $T_{z=z_{\mathrm{CO}}}$ and the observed brightness temperature. Panels (j), (k), and (l) show the extracted perturbations of the perturbed $T_{z=z_{\mathrm{CO}}}$ using the same filtering technique mentioned in Section \ref{M32}.}
    \label{fig:pt-pf-t}
\end{figure*}

As $T_{z=z_{\mathrm{CO}}}$ was fixed to $T_{\mathrm{B}}$ when the perturbations in $n(\mathrm{H_2})_{z=z_{\mathrm{CO}}}$ and $z_{\mathrm{CO}}$ were inferred, it is important to also test how large the perturbations in the gas temperature would need to be to account for the observed perturbations in $v_{\mathrm{rot}}$. Similar to $n(\mathrm{H_2})_{z=z_{\mathrm{CO}}}$ and $z_{\mathrm{CO}}$, we multiply the Gaussian perturbations, $\delta (r)$, to $T_{z=z_{\mathrm{CO}}}$, and find best-fit values for the perturbed $T_{z=z_{\mathrm{CO}}}$ that are able to fully cover the perturbations in $v_{\mathrm{rot}}$. The result is shown in Figure \ref{fig:pt-pf-t}, following the same structure of Figure \ref{fig:pt-pf-n}. We find that the required perturbations can be up to $10\%$ relative to their background values, which are much larger than the observed deviations in $T_\mathrm{B}$ profile, suggesting that perturbations in the gas temperature would only be a minor contribution to the observed velocity deviations. Furthermore, the perturbations are in radial locations that are inconsistent with the $T_\mathrm{B}$ profiles. Hence, we conclude that the deviations in $v_{\mathrm{rot}}$ are likely dominated by perturbations in $n(\mathrm{H_2})_{z=z_{\mathrm{CO}}}$ and $z_{\mathrm{CO}}$, rather than $T_{z=z_{\mathrm{CO}}}$.

\section{Discussion} \label{sec:discussion}

\subsection{Meridional flows} \label{sec:meridional flows}
As shown in Figure \ref{fig:32}, we have detected a collapsing gas flow around 125 au (indicated by a black dotted box), and tentatively a second flow at 210 au, suggestive of a meridional flow. Meridional flows were previously detected in HD~163296 \citep{2019Natur.574..378T}, where they were shown to have larger velocities in the atmospheric layers than in the midplane - consistent with what is observed for HD~169142. Such velocity structures require a large depletion in the gas surface density (e.g., \citealt{2014Icar..232..266M,2014ApJ...782...65S}), with several studies (e.g., \citealt{1984ApJ...285..818P}) showing that such depletion can be created by planet-disk interactions, particularly for giant planets. However, other scenarios -- unrelated to planets, such as the magneto-rotational instability (MRI; \citealt{2015A&A...574A..68F}) or the vertical shear instability (VSI; \citealt{2017ApJ...835..230F,2019A&A...625A.108R}), may also cause depletion of gas surface density and potentially form similar velocity structures, and cannot be ruled out as explanations for such collapsing flow. Future simulations can be proposed to help discriminate different origins of such flow.

\subsection{Vertical disk model at the gap center} \label{sec:vt model}

\begin{figure}
    \centering
    \includegraphics[width=\columnwidth]{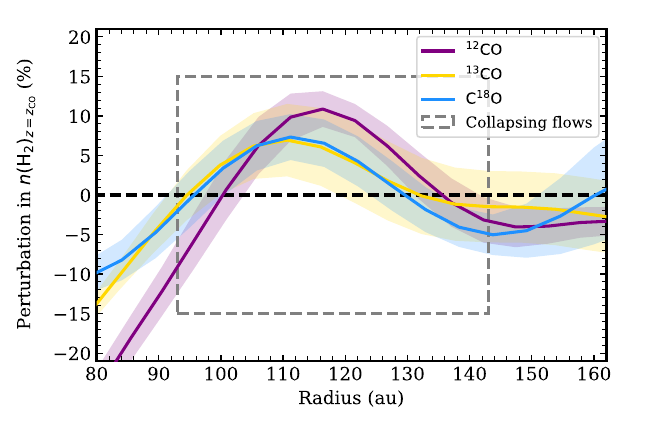}
    \caption{A summary of the radial profiles of the estimated perturbation in $n(\mathrm{H_2})_{z=z_{\mathrm{CO}}}$ traced by $^{12} \mathrm{CO \ (2-1)}$, $^{13} \mathrm{CO \ (2-1)}$, and $\mathrm{C}^{18} \mathrm{O \ (2-1)}$ emissions. The shaded area shows the error for each curve. The grey dotted box indicates the radial range of the collapsing flows found in the $v_{\mathrm{rad}}$ and $v_{\mathrm{z}}$ structure in Figure \ref{fig:32}.}
    \label{fig:pt-n-sum}
\end{figure}

Based on the changes in rotational velocity, we infer an increase in the gas number density at the same location as the aforementioned meridional flow at 125~au. This is found for all three CO isotopologues, as shown in Figure \ref{fig:pt-n-sum}. An increase in gas density is naively counter to the gas density depletions required for meridional flows. However, \citet{2020A&A...642A.165R} used thermo-chemical models to investigate the impact of dust and gas gaps on the gas properties, finding that a gas gap will lead to complex changes in the expected properties of gas density tracers. For instance, with a shallow gap in the gas surface density, CO emission will trace a deeper (i.e., a lower $z_{\mathrm{CO}}$) region of the disk, but probe hotter (i.e., a higher $T_{z=z_{\mathrm{CO}}}$) and denser (i.e., a higher $n(\mathrm{H_2})_{z=z_{\mathrm{CO}}}$) gas, despite the drop in surface density, similar to what is inferred for HD~169142. With a very deep gap, the CO emission was found to trace a less dense (i.e., a lower $n(\mathrm{H_2})_{z=z_{\mathrm{CO}}}$) region, as would naively be expected for a surface density depletion. This modeling suggests that the relative changes in emission height traced by a molecule and the background gas density can act as a good proxy for the total gas surface density depletion.

We also try to look for structures around 125~au in the Moment 0 maps and their respective profiles in Fig.~\ref{fig:ov-gas1}, but no obvious substructures are detectable. This is because the kinematic method probes changes in the rotational velocity that is the derivative of the gas pressure. As such, the spatial scales of the features are bigger for kinematic features than in the intensity \citep{2018ApJ...860L..12T}. Future higher angular resolution observations can be proposed to search for gaps in the Moment 0 maps around 125~au. Note, however, that the line emission is likely optically thick, such that the structures inferred from the gas pressure may not be directly reflected in the intensity of the molecular emission as the kinematic technique leverages spatial variations in the emission distribution, rather than intensity variations. This means even if we have a higher resolution observation in the future, such structures may only be shown in the kinematics instead of emission maps.

To test this hypothesis, and to reconcile the apparent increase in gas number density (inferred by changes in $v_{\mathrm{rot}}$) at a surface density minimum (suggested by the meridional flows), we develop an 1D analytical model of the disk vertical structure, to explore how changes in the gas surface density are related to the changes in $z_{\mathrm{CO}}$, $T_{z=z_{\mathrm{CO}}}$ and $n(\mathrm{H_2})_{z=z_{\mathrm{CO}}}$ inferred through $v_{\mathrm {rot}}$ measurements.

\subsubsection{Model setup} \label{M521}
\begin{figure}
    \centering
    \includegraphics[width=\columnwidth]{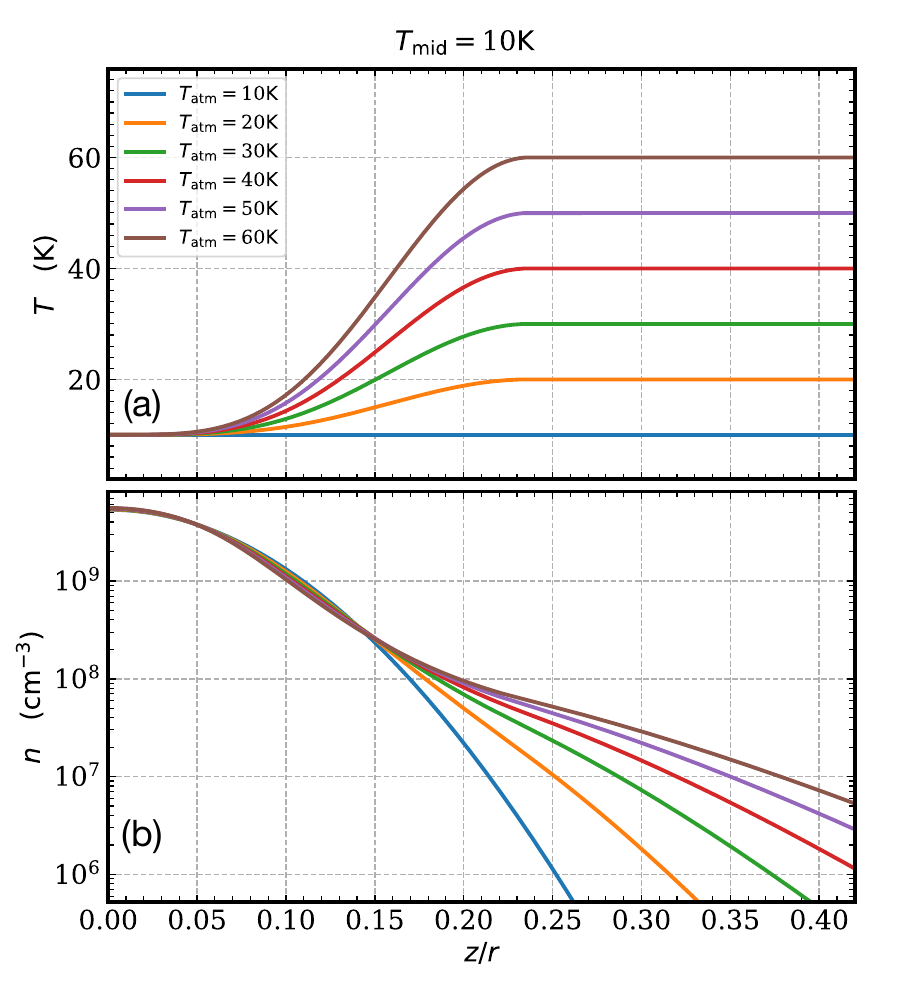}
    \caption{Vertical temperature structure, $T(z)$, and density structure, $n(z)$, at the gap center are shown in panels (a) and (b), respectively. $T(z)$ is calculated from Equation (\ref{eq:vt-temp}). We fix $T_{\mathrm {mid}} = 10 \ \mathrm{K}$, and vary $T_{\mathrm {atm}}$ from 10 K to 60 K in steps of 10~K. $n(z)$ is calculated by numerically solving Equation (\ref{eq:vt-den}) for each different $T(z)$ in panel (a).}
    \label{fig:vt-td}
\end{figure}

We consider a vertical column that is in hydrostatic equilibrium with a vertical temperature gradient and a fiducial surface density of $\Sigma_0$. Following \citet{2014ApJ...788...59W}, we model the vertical temperature structure as

\begin{equation} \label{eq:vt-temp}
T(z)=\left\{\begin{array}{ll}
T_{\mathrm {mid}}+\left(T_{\mathrm {atm}}-T_{\mathrm {mid}}\right)\left[\sin \left(\frac{\pi z}{2 z_{q}}\right)\right]^{4} & \mathrm { if } z<z_{q} \\
T_{\mathrm {atm}} & \mathrm { if } z \geqslant z_{q},
\end{array}\right.
\end{equation}

where $T_{\mathrm {mid}}$ is the midplane temperature, $T_{\mathrm {atm}}$ is the atmospheric temperature, and $z_{q}$ represents the transition to the disk atmosphere where $T(z)=T_{\mathrm {atm}}$. Following \citet{2014ApJ...788...59W}, we fix $z_{q}=4H_{\mathrm{p}}$, where $H_{\mathrm{p}}$ is the midplane pressure scale height, given by,

\begin{equation}
H_{\mathrm{p}} = \sqrt{\frac{k T_{\mathrm{mid}} r^{3}}{G M_{\mathrm{star}} \mu m_{\mathrm{H}}}},
\end{equation}

and $m_{\mathrm{H}}$ is the mass of a hydrogen atom, $\mu = 2.37$ is the mean molecular weight of the gas in the disk. We fix $T_{\mathrm {mid}} = 10 \ \mathrm{K}$ as this is typical of midplane temperatures \citep{2016A&A...586L...1G}, and vary $T_{\mathrm {atm}}$ from 10 K to 60 K in steps of 10~K to cover the range of temperatures likely traced by CO isotopologue emissions based on their brightness temperatures (shown in Fig.~\ref{fig:ov-gas1}c and d). We show these vertical temperature profiles in panel (a) of Figure \ref{fig:vt-td}.

Given the vertical temperature structure, $T(z)$, the vertical density structure can be calculated by solving for hydrostatic equilibrium:

\begin{equation} \label{eq:vt-den}
\frac{\partial \ln \rho}{\partial z}=-\left[\left(\frac{G M_{\text {star }} z}{\left(r^{2}+z^{2}\right)^{3 / 2}}\right)\left(\frac{\mu m_{\mathrm{H}}}{k T}\right)+\frac{\partial \ln T}{\partial z}\right].
\end{equation}

This is solved numerically to yield the vertical density structure, $n(z)$ ($n=\rho / m_{\mathrm{gas}}$), for each different $T(z)$ considered. We adopt a surface density of $5~\mathrm{g~cm^{-2}}$, typical of disks with large masses as used by \citet{2014ApJ...788...59W}. The resulting vertical density structures are shown in Figure \ref{fig:vt-td}b. 

We include a simple model for the CO emission surface. CO is assumed to be present in large abundance when there is a large enough column density of material above it to sufficiently shield it from dissociating UV radiation \citep[$N_{\mathrm{shield}}$;][]{2009A&A...503..323V}. Below this height, we assume that the abundance of CO is great enough that the emission becomes optically thick as soon as there is enough CO for self-shielding. The emission height of CO, $z_{\mathrm{CO}}$, is therefore the height where the gas column density, integrated from the surface of the disk down towards the midplane, reaches $N_{\mathrm{shield}}$. Thus, $z_{\mathrm{CO}}$ satisfies

\begin{equation}
N_{\mathrm{shield}} = \int_{z_{\mathrm{CO}}}^{\infty} \rho(z) \  {\mathrm d}z.
\label{eq:nshield}
\end{equation} 

In typical ISM conditions, $N_{\mathrm{shield}} = 1.3 \times 10^{21}~{\mathrm{H_2}}~{\rm cm^{-2}}$ \citep{2009A&A...503..323V, 2011ApJ...740...84Q} for $^{12} \mathrm{CO}$. $^{13} \mathrm{CO}$ and $\mathrm{C}^{18} \mathrm{O}$ require a larger $N_{\mathrm{shield}}$ value as they are less abundant than $^{12} \mathrm{CO}$, and so the self-shielding process is less efficient. To account for this in our model, we adopt different $N_{\mathrm{shield}}$ to represent different CO isotopologues. We describe $N_{\mathrm{shield}}$ as $N_{\mathrm{shield}}=k N_{\mathrm 0}$, where $N_{\mathrm{0}}=1.3 \times 10^{21}~{\mathrm H_2}~{\mathrm{cm^{-2}}}$, and $k$ is varied from 1 to 5. Following \citet{2009A&A...503..323V}, we find comparable photodissociation rates to $^{12}$CO ($k = 1$) for $^{13}$CO and C$^{18}$O when $k \approx$ 5 for disk-like conditions (ignoring the influence of dust-shielding). As such, the adopted range of $k$ values provides a reasonable model to reproduce the different vertical layers traced by different molecular species.

With $z_{\mathrm{CO}}$ to hand, and the temperature and density structures shown in Figure \ref{fig:vt-td}, we can calculate the gas temperature and the number density traced by the CO emission, $T_{z=z_{\mathrm{CO}}}$ and $n(\mathrm{H_2})_{z=z_{\mathrm{CO}}}$, respectively. We do not consider $v_{\mathrm {rot}}$ given that it depends on the \emph{radial} pressure gradient, requiring a 2D model instead of 1D model. Moving to a 2D model would require extra assumptions about the gap width and radial dependence of the gas properties, substantially complicating the interpretation.

\subsubsection{Influence of depletion in the surface density} \label{M522}
We parameterize the surface density depletion through a parameter $\delta$, such that the perturbed surface density is given by $\delta \times \Sigma_0$. $\delta$ varies between 1, for no perturbation, and 0, representing a completely empty gap. We vary $\delta$ to simulate different gap depths, and explore how the emission height of CO changes, and thus the temperature, $T_{z=z_{\mathrm{CO}}}$, and number density, $n(\mathrm{H_2})_{z=z_{\mathrm{CO}}}$, traced. We do not consider any change in the physical structure due to changes in the irradiation of the gap, i.e., a change in the temperature structure of the gap, which are discussed in some studies (e.g., \citealt{2017ApJ...835..228T}, \citealt{2018A&A...612A.104F}, \citealt{2018ApJ...867L..14V}). The simplifying choice of the gas temperature was made because the way the gas temperature reacts when a gap is opened is highly dependent on several factors, such as the local gas-to-dust ratio and the gap depth that really require an extensive thermo-chemical model to resolve (see \citealt{2020A&A...642A.165R}, for different ‘responses’ within the same disk). Under this assumption, any change in the pressure will be entirely due to changes in the density structure of the disk as the temperature structure remains fixed.

\begin{figure*} 
    \centering
    \includegraphics[width=0.9\textwidth]{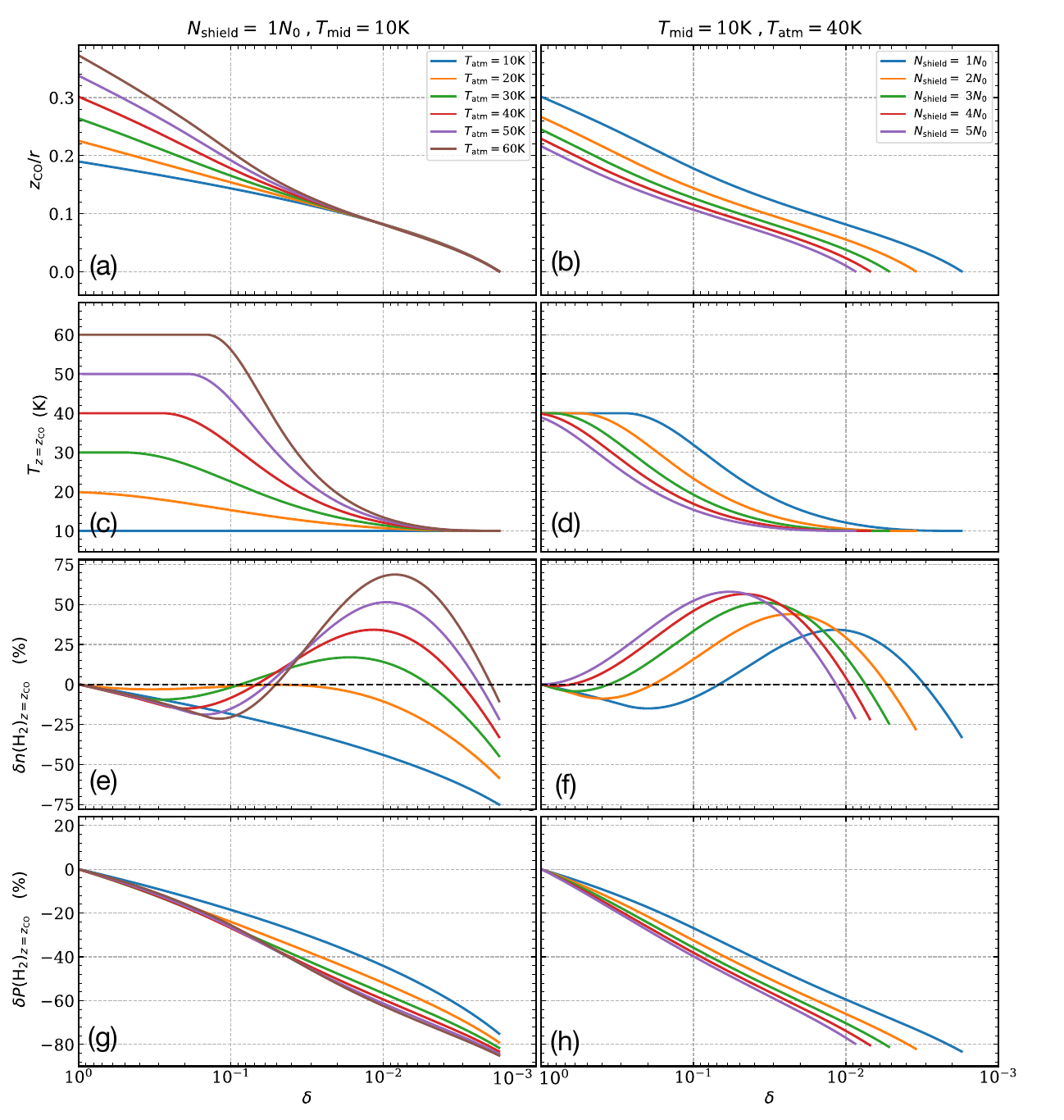}
    \caption{CO-traced gas emission height, $z_{\mathrm{CO}}$, temperature, $T_{z=z_{\mathrm{CO}}}$, number density, $n(\mathrm{H_2})_{z=z_{\mathrm{CO}}}$, and gas pressure, $P(\mathrm{H_2})_{z=z_{\mathrm{CO}}}$, which are the functions of surface density depletion, $\delta$. The left column compares different temperature structures. We do not consider any change in the physical structure due to changes in the irradiation of the gap, i.e., a change in the temperature structure of the gap. We fix $N_{\mathrm{shield}}=N_0$ and $T_{\mathrm {mid}} = 10~\mathrm{K}$, and vary $T_{\mathrm {atm}}$ between 10~K and 60~K with steps of 10~K. The right column shows a comparison between different values of $N_{\mathrm{shield}}$, representing different isotopologues. We fix $T_{\mathrm {mid}} = 10~\mathrm{K}$ and $T_{\mathrm {atm}} = 40 \ \mathrm{K}$, and vary $k$ in $N_{\mathrm{shield}}=kN_0$ from 1 to 5. Panels (a) and (b) show $z_{\mathrm{CO}}$, panels (c) and (d) show $T_{z=z_{\mathrm{CO}}}$, panels (e) and (f) show $\delta n(\mathrm{H_2})_{z=z_{\mathrm{CO}}}$, and panels (g) and (h) show $\delta P(\mathrm{H_2})_{z=z_{\mathrm{CO}}}$. Both $\delta n(\mathrm{H_2})_{z=z_{\mathrm{CO}}}$ and $\delta P(\mathrm{H_2})_{z=z_{\mathrm{CO}}}$ are the fractional percentage differences relative to the unperturbed values.}
    \label{fig:vt-sum}
\end{figure*}

We calculate $z_{\mathrm{CO}}$ for different values of $\delta$ shown in panels (a) and (b) of Figure \ref{fig:vt-sum}. In panel (a), we fix $N_{\mathrm{shield}} = N_0$, and vary the temperature structure. Conversely, panel (b) shows a comparison for different $N_{\mathrm{shield}}$ values, representing different isotopologues, while the temperature structure is held constant with $T_{\mathrm {mid}} = 10~\mathrm{K}$ and $T_{\mathrm {atm}} = 40 \ \mathrm{K}$. For this panel we vary $k$ from 1 to 5. We find that for all scenarios, $z_{\mathrm{CO}}/r$ decreases with decreasing $\delta$, i.e., a deeper gap. As expected, all surface density depletions result in a drop in emission height. As shown in panel (a), for the same $T_{\mathrm {mid}}$, but higher $T_{\mathrm {atm}}$, the $z_{\mathrm{CO}}/r$ decreases at a faster rate for a given change in $\Sigma$ for small perturbations, but all curves tend to converge for significant depletions, e.g., after $\delta < 10^{-2}$. This is because after $\delta < 10^{-2}$,  $z_{\mathrm{CO}}/r \le 0.1$ for all $k$, where there is a comparable $n$ structure for all different temperature structures (see Figure \ref{fig:vt-td}b). In panel (b), we see that the change in $z_{\mathrm{CO}}$ with $\delta$ is the same for all $N_{\mathrm{shield}}$ values.

Panels (c) and (d) in Figure \ref{fig:vt-sum} show how $T_{z=z_{\mathrm{CO}}}$ changes for different background temperature structures and different values of $N_{\mathrm{shield}}$, respectively. For all scenarios, $T_{z=z_{\mathrm{CO}}}$ decreases with  a deeper gap (decreasing $\delta$). This is because in our model, $T$ always decreases towards the midplane (see Figure \ref{fig:vt-td}a) and $z_{\mathrm{CO}}/r$ decreases monotonically with $\delta$ (see Figure \ref{fig:vt-sum}a, b). 

Panels (e) and (f) in Figure \ref{fig:vt-sum} show $\delta n(\mathrm{H_2})_{z=z_{\mathrm{CO}}}$ for different temperature structures and different values of $N_{\mathrm{shield}}$, respectively. According to Equation (\ref{eq:4}), the $v_{\mathrm{rot}}$ is dependent on the radial gradient of $n(\mathrm{H_2})_{z=z_{\mathrm{CO}}}$, and so we only want to know how $n(\mathrm{H_2})_{z=z_{\mathrm{CO}}}$ changes relative to its unperturbed value, $n(\mathrm{H_2})_{z=z_{\mathrm{CO}}}(\delta=1)$. Hence we plot the percentage change relative to this unperturbed value. In most scenarios we consider, $\delta n(\mathrm{H_2})_{z=z_{\mathrm{CO}}}$ shows a similar pattern: as we decrease $\delta$ (representing a deeper gap), $\delta n(\mathrm{H_2})_{z=z_{\mathrm{CO}}}$ first decreases for small changes in $\delta$, then increases for moderate perturbations, before finally decreasing again at large perturbations. One exception is that in the vertically isothermal case, as indicated by the blue curve in panel (e), the $\delta n(\mathrm{H_2})_{z=z_{\mathrm{CO}}}$ decreases monotonically with $\delta$ -- the behaviour one naively expects for a gas gap. It should also be noted that the amplitude of the fluctuation of the $\delta n(\mathrm{H_2})_{z=z_{\mathrm{CO}}}$ increases with higher $T_{\mathrm {atm}}$, while at higher $N_{\mathrm{shield}}$ values, the $\delta n(\mathrm{H_2})_{z=z_{\mathrm{CO}}}$ curves have turning points at higher $\delta$ values.

We also show $\delta P(\mathrm{H_2})_{z=z_{\mathrm{CO}}}$ in panels (g) and (h) in Figure \ref{fig:vt-sum}, which is the percentage change in gas pressure relative to the unperturbed value, $P(\mathrm{H_2})_{z=z_{\mathrm{CO}}}(\delta=1)$ assuming an ideal gas. For all scenarios, $\delta P(\mathrm{H_2})_{z=z_{\mathrm{CO}}}$ decreases with a deeper gap (decreasing $\delta$), contrary to the pressure profile inferred from the observations.

With these models, we show that the physical structure of the gap can have non-trivial responses (i.e., molecule emission tracing a larger gas number density despite being in a gap) to changes in the surface density, as previously shown in \citet{2020A&A...642A.165R} through a full thermo-chemical modeling approach. Meanwhile, we note that the simplified model only provides absolute values and not the radial gradient that the derivation of the kinematic structure needs -- a radial temperature structure in the gap is needed to disentangle possible scenarios. As with current observation we cannot spatially resolve any changes in gas temperature (see Fig.~\ref{fig:pt-pf-t}), we are unable to differentiate between different possible scenarios. With future observations at high angular resolution and more extensive chemical modeling, we will be able to leverage the kinematical constraints to better constrain the physical structure of the gap.

\subsection{Interpretations}
We compare the result of the above 1D modeling with the perturbed $n(\mathrm{H_2})_{z=z_{\mathrm{CO}}}$ estimated in Section \ref{sec:model}, and shown in Figure \ref{fig:pt-n-sum}. As the radial pressure gradient is calculated at a constant height, rather than along the emission surface, when the temperature traced by CO emission decreases with an increasing gap depth, the radial temperature gradient remains fixed as the temperature structure has not changed (although a negligible change is expected if $T_{\rm mid}$ and $T_{\rm atm}$ have different radial dependencies as then the background temperature gradient would then be dependent on height). Conversely, as the gap is modeled as a depletion in the gas surface density, the density structure of the entire vertical column does change, resulting in a comparable change in the radial density profile, and thus the radial pressure profile.

Here we use the typical case of $T_{\mathrm {mid}} = 10~\mathrm{K}$, $T_{\mathrm {atm}} = 40 \ \mathrm{K}$, and $N_{\mathrm{shield}}=N_0$ ($k = 1$) in the above 1D model as an example to compare with the perturbed $n(\mathrm{H_2})_{z=z_{\mathrm{CO}}}$ in Figure \ref{fig:pt-n-sum}. We arbitrarily define $1 \ge \delta \geq 0.07$ as a shallow gap, $0.07 \geq \delta \ge 0.003$ as a moderate gap, and $\delta \le 0.003$ as a deep gap. In our model (Fig.~\ref{fig:vt-sum}e, f), we find that for shallow and deep gaps, the emission traces a negative $\delta n(\mathrm{H_2})_{z=z_{\mathrm{CO}}}$, and thus a lower gas number density. While for moderate gaps, the emission traces a positive $\delta n(\mathrm{H_2})_{z=z_{\mathrm{CO}}}$, correspondingly higher gas number densities. This is because, with a depletion in the surface density, $z_{\mathrm{CO}}$ decreases, indicating that we are tracing a denser region according to the vertical density structure (see Fig.~\ref{fig:vt-td}b). The tracing of a denser region competes with the decrease of the surface density, reducing the gas volume density at a given height. Therefore, the emission will trace a perturbation ranging from a positive $\delta n(\mathrm{H_2})_{z=z_{\mathrm{CO}}}$ to a negative $\delta n(\mathrm{H_2})_{z=z_{\mathrm{CO}}}$ depending on the results of this competition. This is consistent with the findings in \citet{2020A&A...642A.165R}, where the authors demonstrate too the possibility that the CO emission will trace a denser region of the disk (i.e., a higher $n(\mathrm{H_2})_{z=z_{\mathrm{CO}}}$) despite the drop in surface density. 

The inferred structure at $r=125~{\mathrm{au}}$, where a surface density minimum is required based on the $v_{\mathrm{rad}}$ and $v_{\mathrm z}$ structure, but CO isotopologue emission traces higher gas volume densities, is consistent with our modeling for a moderate gap, i.e., $0.07 \geq \delta \ge 0.003$. Note that when interpreting these findings, we want to use the described model only as a guide as the true physical structure of the gaps will be far more complex. For example, while $v_{\mathrm{rot}}$ is sensitive to both $n(\mathrm{H_2})_{z=z_{\mathrm{CO}}}$ and $z_{\mathrm{CO}}$, we have considered $n(\mathrm{H_2})_{z=z_{\mathrm{CO}}}$ separately in Section~\ref{sec:model}, meaning that the estimated scale of $\delta$ here is only poorly constrained. As the disk around HD~169142 is observed from an almost face-on orientation, any information (e.g., emission height) about the vertical disk structure cannot be extracted from the observation. Future studies can repeat this analysis for disks at moderate inclination, where the emission height can be directly measured and thus used to break some of the degeneracy between the $n(\mathrm{H_2})_{z=z_{\mathrm{CO}}}$ and $z_{\mathrm{CO}}$. 

To provide some representative numbers to demonstrate that a planet-opened gap is a plausible scenario, and is consistent with the observed features, we predict the mass of the potential planet opening this gap based on the estimated gap depth, $0.07 \geq \delta \ge 0.003$. Following the method outlined in \citet{2015ApJ...806L..15K,2018ApJ...869L..47Z,2019ApJ...884..142G}, we estimate that the mass of the planet is between 0.5 to $2.5~{M_\mathrm{Jup}}$ assuming the disk viscosity $\alpha = 1 \times 10^{-3}$ \citep{2015ApJ...813...99F,2017ApJ...843..150F,2020ApJ...895..109F,2016A&A...592A..49T,2018ApJ...868..113T}. Assuming a lower $\alpha$, i.e., $1 \times 10^{-4}$, the estimated planet mass can be as low as 0.15~$M_{\mathrm {Jup}}$. Estimating the planet mass based on the width of this gap or the velocity perturbation \citep{2019ApJ...884..142G} results in a mass $\sim 1~{M_\mathrm{Jup}}$.

\section{Conclusion} \label{sec:conclusion}
We present an analysis of archival ALMA observations of $^{12} \mathrm{CO \ (2-1)}$, $^{13} \mathrm{CO \ (2-1)}$, and $\mathrm{C}^{18} \mathrm{O \ (2-1)}$ emission from the disk around HD~169142 to search for large-scale kinematic structures. We extract 3D kinematical information at different vertical layers within the disk traced by the different CO isotopologues. We use analytical models to explore the dynamical and physical structure of the disk that would be consistent with the measured velocity profiles. The key findings are as follows:

\begin{enumerate}
\item We have extracted 3D velocity structure with the method following \citet{2018ApJ...860L..12T, 2018ApJ...868..113T, 2019Natur.574..378T}. We identify collapsing gas flows around 125~au, and tentatively 210 au. At 125~au, the flow appears to be continuous, traced by all three CO isotopogoluges. These features are interpreted as meridional flows, bearing a striking resemblance to those modeled for the HD 163296 disk in \citet{2019Natur.574..378T}.

\item We observe deviations up to $1\%$ in the $v_{\mathrm{rot}}$ profiles, relative to the background rotation. We  estimate the perturbations required to drive such deviations if they were to be from either changes in the gas number density, gas temperature, or the emission height. We find that the required perturbations can be up to $10\%$ relative to a smooth background for all three properties. As these variations are much higher than those observed in the $T_\mathrm{B}$ profiles, we argue the perturbations are likely caused mainly by changes in the gas number density or the emission height.

\item We develop a simplified analytical model of a gap, and find that changes in the gas surface density can have non-trivial impacts on the physical structure of the gap where the molecular emission is tracing. These results are consistent with a full thermo-chemical modeling approach presented in \citet{2020A&A...642A.165R}. We also show possibility of inferring the mass of the planet responsible for opening the gap from such responses. While we cannot readily explain the kinematic deviation with this simplified model, these sort of observations provide a tool to access the temperature and density profiles within the gap and that when combined with extensive thermo-chemical modeling, provide the best opportunity to characterize the physical structure of a gap.
\end{enumerate}

In sum, this work demonstrates the ability to map the 3D kinematical structure of protoplanetary disks, and serves as a foundation for future studies to characterize the physical conditions around gap edges.

\acknowledgments
The authors thank the anonymous referee for constructive comments that improved this work. This paper makes use of the following ALMA data: 2013.1.00592.S, 2015.1.00490.S and 2015.1.01301.S. ALMA is a partnership of ESO (representing its member states), NSF (USA) and NINS (Japan), together with NRC (Canada), NSC and ASIAA (Taiwan), and KASI (Republic of Korea), in cooperation with the Republic of Chile. The Joint ALMA Observatory is operated by ESO, AUI/NRAO and NAOJ. The National Radio Astronomy Observatory is a facility of the National Science Foundation operated under cooperative agreement by Associated Universities, Inc. R.T. acknowledges support from the Smithsonian Institution as a Submillimeter Array (SMA) Fellow. J.B. acknowledges support by NASA through the NASA Hubble Fellowship grant \#HST-HF2-51427.001-A awarded by the Space Telescope Science Institute, which is operated by the  Association of Universities for Research in Astronomy, Incorporated, under NASA contract NAS5-26555.

\facilities{ALMA}
\software{CASA \citep[5.8.0;][]{McMullin_ea_2007}, GoFish \citep{GoFish}, bettermoments \citep{2018RNAAS...2c.173T}, eddy \citep{2019JOSS....4.1220T}, dynesty \citep{2020MNRAS.493.3132S}}

\newpage
\appendix
\section{Different methods for calculating the velocity maps} \label{apd:1}
\begin{figure}
    \centering
    \includegraphics[width=\textwidth]{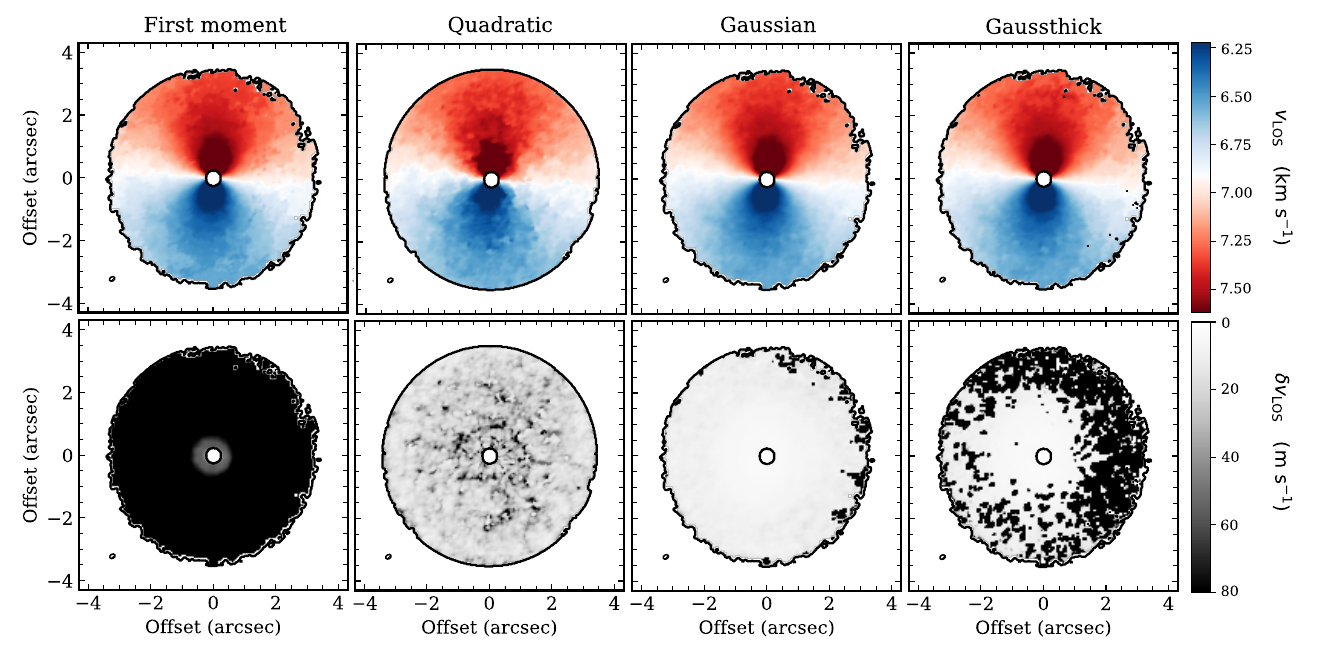}
    \caption{Comparison between different methods for calculating the velocity maps. The first row shows the velocity maps calculated by the four methods: the intensity-weighted average velocity (a \texttt{first moment} map), \texttt{quadratic}, \texttt{gaussian}, and \texttt{gaussthick}. The second row shows the associated uncertainties. (Use $^{12} \mathrm{CO \ (2-1)}$ as an example.)}
    \label{fig:vcomp}
\end{figure}

We have tested different methods for calculating the line-of-sight velocity, $v_{\mathrm{LOS}}$, using the \texttt{bettermoments} package. These methods include: the intensity-weighted average velocity, a \texttt{first moment} map; \texttt{quadratic}, fitting a quadratic function to the brightest pixels of the spectrum and its two neighboring pixels; \texttt{gaussian}, fitting a Gaussian function to each pixel; and \texttt{gaussthick}, similar to \texttt{gaussian}, but including an approximation of an optically thick line core \citep[e.g.,][]{2016A&A...591A.104H}. We show a comparison of the performance of these methods in Figure \ref{fig:vcomp}, using velocities traced by $^{12} \mathrm{CO \ (2-1)}$ as an example. The first row shows the velocity maps calculated by the four methods, while the second row shows the associated uncertainties.

\begin{figure}
    \centering
    \includegraphics[width=1.0\textwidth]{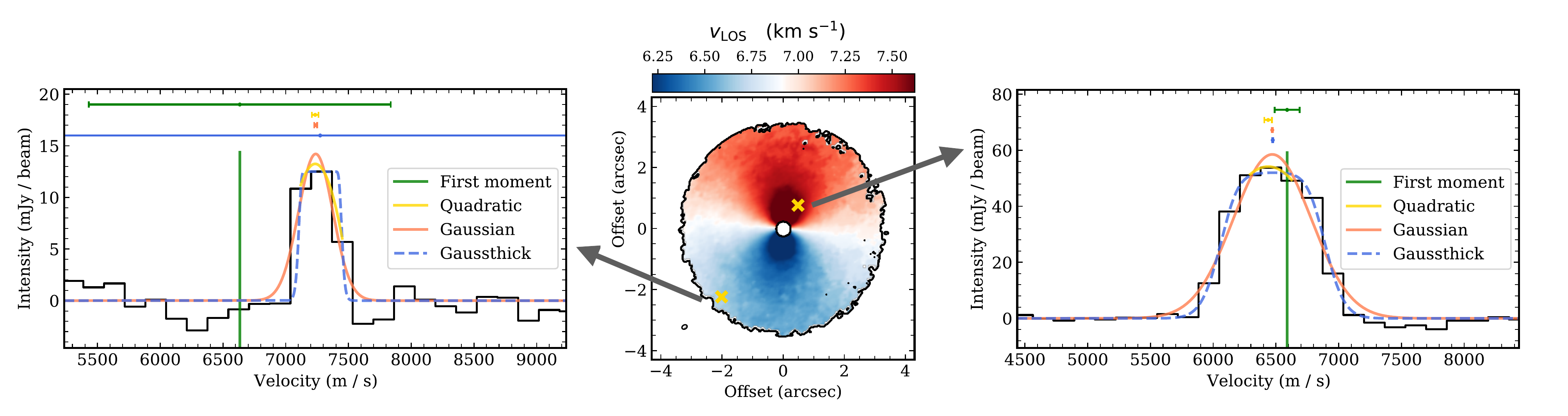}
    \caption{Zoom-in comparison of difference among methods for inferring the line center $v_{\mathrm{LOS}}$ of the spectra using $^{12}$CO (2-1) as an example. The locations from where the spectra in the left and right panels are extracted are labeled by yellow crosses on the velocity map in the middle panel. The left and right panels show how the different functional forms fit those spectra. The dots and the error bars in the upper part of these panels show the inferred $v_{\mathrm{LOS}}$ and their statistical uncertainties from the four methods.}
    \label{fig:vcomp-zoom}
\end{figure}

We also zoom into a few example pixels to compare in details the difference among methods for inferring the line center, $v_{\mathrm{LOS}}$, in Figure \ref{fig:vcomp-zoom}. The examples pixels are labeled by yellow crosses in the example velocity map in the middle panel. The left and right panels show how the different functional forms fit those spectra. The dots and the error bars in the upper part of these panels show the inferred $v_{\mathrm{LOS}}$ and associated statistical uncertainties from the four methods.

From the comparisons, we find that the \texttt{gaussian} method fits the best to the spectra, as well as provides the lowest uncertainties around both the center and the edge of the velocity map, while the \texttt{first moment} method returns the largest uncertainties. The \texttt{gaussthick} method also works well around the center of the map, but gets significantly worse when it comes close to the edge of the velocity map. As described in \citet{2018RNAAS...2c.173T}, the \texttt{quadratic} approach yields a robust measure of the line center even for asymmetric line profiles. If the underlying profile is close to a Gaussian, as is the case for HD~169142, then fitting an analytical form, such as the \texttt{gaussian} or \texttt{gaussthick} methods, will always yield more precise measures. For the specific case of the data used for this paper, the spectra are too coarse to find a better constraint with the \texttt{gaussthick} method, and large covariances between the optical depth and the line center result in only poor constraints.

\section{Radial profiles for 3D velocity components} \label{apd:2}
\subsection{Enhancement on the velocity profiles} \label{apd:21}

\begin{figure}
    \centering
    \includegraphics[width=1.0\textwidth]{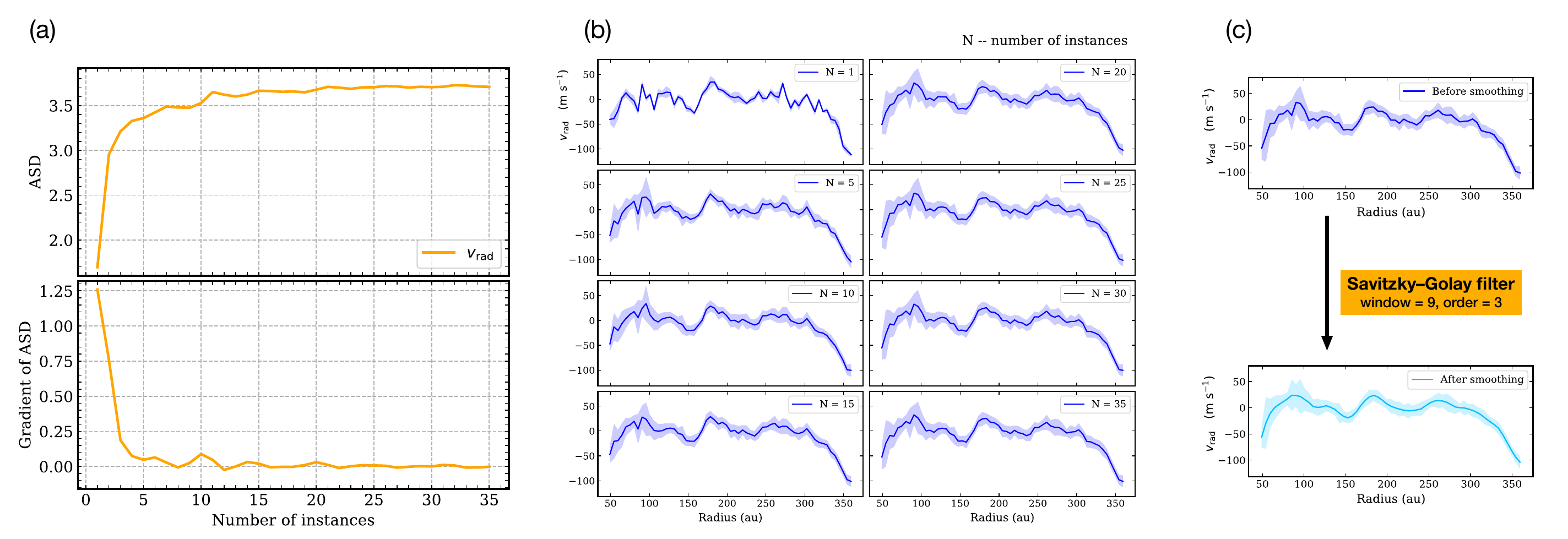}
    \caption{Panel (a) shows the ASD and the gradient of the ASD versus the number of instances. Panel (b) shows examples of how $v_{\mathrm{rad}}$ profiles change with more instances averaged corresponding to panel (a). Panel (c) shows a comparison between before and after implementing the Savitzky-Golay filter with a window of 9 data points ($\sim$ 47 au) and polynomial order of 3. (Use $^{12} \mathrm{CO \ (2-1)}$ as an example.)}
    \label{fig:6}
\end{figure}

We use the method described in Section \ref{M31} to extract the profiles of 3D velocity components: rotational velocity, $v_{\mathrm{rot}}$, radial velocity, $v_{\mathrm{rad}}$, and vertical velocity, $v_{\mathrm{z}}$. We split the data into annuli, and run MCMC with 32 walkers, 500 burn-in steps and 500 steps on each annulus to explore the posterior distributions of model parameters of $v_{\mathrm{rot}}$ and $v_{\mathrm{rad}}$, in order to estimate the best-fit values. For each annulus, spatially-independent pixels are randomly selected. After one instance of the fitting procedure, we find that there is a great amount of high frequency noise mixed in the profiles, which interferes with our search for the real perturbations. To better characterize the range of velocity profiles consistent with the data, we run multiple instances of the fitting procedure, each time selecting a random set of independent pixels, and taking the mean and standard deviation of the ensemble of samples as the best-fit value and associated uncertainty (as in \citealt{2019A&A...625A.118K}). To determine how many instances are needed, we calculate the average standard deviation (ASD), which is estimated by calculating the standard deviations of all samples at each radius, then taking the average across all radii. Using $v_{\mathrm{rad}}$ traced by $^{12} \mathrm{CO \ (2-1)}$ as an example, we show the ASD and the gradient of the ASD versus the number of instances in Figure \ref{fig:6}a. Given that the gradient of ASD is almost flat after around 25 instances, we therefore conclude that 25 instances of the fitting procedure are sufficient, that adding more instances hardly further improves the velocity profiles (i.e., suppress the high frequency noise). We confirm this further from Figure \ref{fig:6}b, which shows examples of how $v_{\mathrm{rad}}$ profiles change with more instances averaged. We see that there are apparent improvements before 25 instances, but no noticeable changes afterwards. 

The final step of improving the profiles is to filter out the still remaining high-frequency noise (i.e., see Figure \ref{fig:6}b, N = 25). We find that a Savitzky-Golay filter with a window of 9 data points ($\sim$ 47 au) and polynomial order of 3 performs best for our case, as it would not introduce or eliminate perturbations in the profiles. An example ($v_{\mathrm{rad}}$ traced by $^{12} \mathrm{CO \ (2-1)}$) is shown Figure \ref{fig:6}c.

\subsection{Modeling the background rotation} \label{apd:22}

In order to identify perturbations in $v_{\mathrm{rot}}$, a background model must be subtracted. Due to the background pressure gradient and any potential change in the emission surface, a simple Keplerian model does not suffice \citep[e.g.,][]{2019Natur.574..378T}. In this work, we adopt a new approach, employing a Butterworth low-pass filter to extract a background model based on the low-frequency components of the $v_{\mathrm{rot}}$ curve. In this Appendix, we compare this approach with analytical background models, and discuss the choice of filter properties.

\begin{figure}
    \centering
    \includegraphics[width=1.0\textwidth]{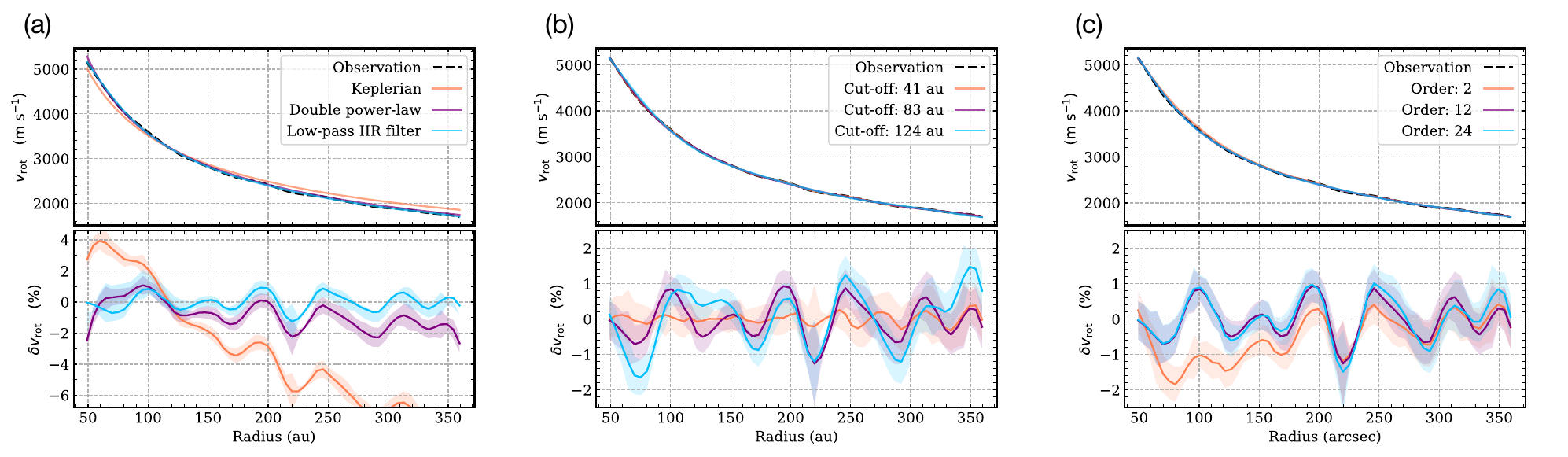}
    \caption{Comparisons among different methods for extracting $\delta v_{\mathrm{rot}}$ profiles, using $v_{\mathrm{rot}}$ traced by $^{12} \mathrm{CO \ (2-1)}$ as examples. The upper figures show the fittings to the $v_{\mathrm{rot}}$ baselines, and the lower figures show the extracted $\delta v_{\mathrm{rot}}$ profiles. Panel (a) shows a comparison among different baseline models: a Keplerian model, a double power-law model, and a non-parametric model extracted by IIR Butterworth low-pass filter. Panels (b) and (c) show comparisons between different cut-offs and orders, respectively, of the implemented IIR Butterworth low-pass filter.}
    \label{fig:ex_perb_comp}
\end{figure}

In Fig.~\ref{fig:ex_perb_comp}a, we compare different baseline models: a Keplerian model, a double power-law model, and a non-parametric model extracted by IIR Butterworth low-pass filter. The upper figure shows the baseline models with the observed $v_{\mathrm{rot}}$, while the lower figure shows the inferred deviations, $\delta v_{\mathrm{rot}}$. We find that the Butterworth filter method performs the best, as there is the smallest overall gradient in the extracted $\delta v_{\mathrm{rot}}$ profile. In panel (b), we compare different cut-offs of the filter. We find that a cut-off smaller than 16 samples (i.e., 8 samples, $\sim$ 41 au) suppresses the real perturbations in $v_{\mathrm{rot}}$, while a cut-off bigger than 16 samples (i.e., 24 samples, $\sim$124 au) leaves apparent overall gradients in the $\delta v_{\mathrm{rot}}$ profiles. In panel (c), we compare different orders of the filter. We find that an order lower than 12 (i.e., 2) leaves remaining baseline in the $\delta v_{\mathrm{rot}}$ profile, as the cutting-off of the low-frequency signal is not steep enough, while an order higher than 12 (i.e., 24) presents a very similar result as the order of 12.

\bibliography{rf1}{}
\bibliographystyle{aasjournal}

\end{document}